\documentclass[12pt]{article}

\usepackage[utf8]{inputenc} 

\usepackage[margin=2cm]{geometry}

\usepackage{graphicx} 

\usepackage{cite}
\usepackage{lineno}

\usepackage{booktabs} 
\usepackage{array} 
\usepackage{paralist} 
\usepackage{verbatim} 
\usepackage{subfigure} 

\usepackage{hyperref}
\usepackage{amsmath}
\usepackage{amsthm}
\usepackage{amssymb}
\usepackage{authblk}
\title{Differential Parametric Formalism for the Evolution of Gaussian
States: Nonunitary Evolution and Invariant States}

\author[1,2]{Julio A. L\'opez-Sald\'ivar}
\author[3]{Margarita A. Man'ko} 
\author[2,3,4]{Vladimir I. Man'ko}

\affil[1]{Instituto de Ciencias Nucleares, Universidad Nacional Aut\'onoma de M\'exico,
Apdo. Postal 70-543, Ciudad de México 04510, M\'exico}
\affil[2]{Moscow Institute of Physics and Technology, Institutskii per. 9,
Dolgoprudnyi, Moscow~Region 141700, Russia}
\affil[3]{Lebedev Physical Institute,  Leninskii Prospect 53,
Moscow,
119991,
Russia}
\affil[4]{
 Department of Physics, Tomsk State University, Lenin Avenue 36, Tomsk 634050, Russia
}
\date{julio.lopez.8303@gmail.com}

\begin{document}

\maketitle

\begin{abstract}In a differential approach elaborated, we study the
evolution of the parameters of Gaussian, mixed, continuous variable density
matrices, whose dynamics are given by Hermitian Hamiltonians expressed as
quadratic forms of the position and momentum operators or quadrature
components. Specifically, we obtain in generic form the differential equations
for the covariance matrix, the mean values, and the density matrix parameters
of a multipartite Gaussian state, unitarily evolving according to a
Hamiltonian $\hat{H}$. We also present the corresponding differential
equations which describe the nonunitary evolution of the subsystems. The
resulting nonlinear equations are used to solve the dynamics of the system
instead of the Schr\"odinger equation. The formalism elaborated allows us to
define new specific invariant and quasi-invariant states, as well as states
with invariant covariance matrices, i.e., states were only the mean values
evolve according to the classical Hamilton equations. By using density
matrices in the position and in the tomographic-probability representations,
we study examples of these properties. As examples, we present novel invariant
states for the two-mode frequency converter and quasi-invariant states for the
bipartite parametric amplifier.
\end{abstract}


\section{Introduction}
The study of Gaussian states has been of an essential interest in the last
decades. This type of states, associated to classical random fields, were
considered as a possibility to connect covariance matrices of the states as
quantum density matrices and, with this definition, to study the
quantum--classical relation of randomness with the quantization
procedure~\cite{Khrennikov-book,KhrenFP}. The problems of new developments of
foundations of quantum mechanics and applications of new results in quantum
information and quantum probabilities, as well as in areas like mathematical
finance and economics, attract the attention of the researchers; they are
intensely discussed in the literature~\cite{khren1,khren2,khren3}. The
important role in this development is played by discussing the problems which
appeared from the very beginning of quantum mechanics, like the notion of
quantum system states and interpretation of the states associated in
conventional formulation of quantum mechanics with Hilbert space vectors and
density operators, using the quasiprobability distributions and the
probability distributions containing the complete information on quantum
states. There exists increasing interest to quantum foundations since a deeper
understanding the essence and formalism of quantum theory is needed for the
development of quantum technologies and possibilities to extend the
applications of quantum formalism in physics to all other areas of science
like economy, finance, and social disciplines.

Some examples of Gaussian states of
quantum fields as the coherent, squeezed, and thermal light states are
regularly used in the theoretical and experimental framework of quantum
mechanics, optics, information, and computing. The use of these states in
quantum information has been of particular
importance~\cite{gaussi,adesso,leuchs}. One can list some of the most recent
applications of the use of Gaussian systems -- it has been
demonstrated~\cite{furasek} that it is not possible to distill more
entanglement from a bipartite Gaussian state, using local Gaussian
transformations. In~\cite{paris}, several properties of the purity of Gaussian
states were found. The connection between the symplectic invariants of
bipartite Gaussian states, the von Neumann entropy, and the mutual information
was established in~\cite{serafini}. The extremality of entanglement measures and secret key rates for Gaussian
states was observed in~\cite{wolf}. It was shown~\cite{cerf} that Gaussian
attacks are characterized by an optimum efficiency against eavesdrop
protocols. Quantum illumination of a target using Gaussian light states was
explored by Tan {\it et al}~\cite{shapiro}. A quantum discord for systems of
continuous variables, such as Gaussian states, was implemented
in~\cite{giorda}. In~\cite{arkhipov}, an invariant describing the
nonclassicality in a two-mode Gaussian state was reported. The entanglement of
$m$ modes with other $n$ modes of a Gaussian multipartite system was treated
in~\cite{lami}. The linear response for systems close to steady states under
Gaussian processes was obtained in~\cite{parrondo}. The optimal measurement of
the fidelity of multimode Gaussian states has been studied in~\cite{jeong}. On
the other hand, the study of Gaussian wave packets by nonlinear differential
equations, as the Riccati equation, has been studied in
\cite{cruz1,cruz2,dieter}. Several coherent states have been
defined by the use of quadratic operators \cite{dod_book}. The behavior of
different quantities as covariances in thermal relaxation phenomena has been
also studied in \cite{vale}.

The aim of this work is to present a new way to characterize the dynamics of
Gaussian states using the differential equations for the parameters, which
determine their continuous variable density matrices. The proposed method
makes use of the integrals of motion of such systems, and it can be used to
clarify new aspects of multimode Gaussian quantum states, such as an explicit
form of nonunitary evolution of the states of subsystems and the existence of
invariant states with constant covariance matrices and mean values.

The time evolution of a quantum system was first established by
Schr\"odinger~\cite{Schroed26}. The dynamics of the system given by a
Hamiltonian operator $\hat{H}$ for a pure state $\vert \psi (t)\rangle$ must
follow the Schr\"odinger equation
\[
\hat{H} \vert \psi (t) \rangle = i\hbar \frac{\partial }{\partial t} \vert
\psi (t) \rangle \, ;
\]
this expression corresponds to a second-order differential equation in the
position representation. In the case of an arbitrary state represented by the
density matrix $\hat{\rho}(t)$, which can be not
pure~\cite{Landau,vonNeumann}, the evolution is determined by the von Neumann
equation
\[
i \hbar \frac{\partial}{\partial t}\hat{\rho}(t)=[ \hat{H},
\hat{\rho}(t) ] \, ;
\]
the general solution of this equation is given by the unitary transform
$\hat{U}(t)$, i.e., $\vert \psi (t) \rangle=\hat{U}(t) \vert \psi (0) \rangle$
or $\hat{\rho}(t)=\hat{U}(t) \hat{\rho}(0) \hat{U}^\dagger (t)$, where $\vert
\psi (0) \rangle$ and $\hat{\rho}(0)$ describe the system at time $t=0$. Also
it is a common knowledge that, when the system interacts with an environment,
its dynamics is described by the master equation~\cite{kossa,gori,lind}.

The Gaussian states can be determined by their covariance matrix
$\boldsymbol{\sigma}$ and mean values $\langle q_j \rangle$ and $\langle p_j
\rangle$. This property also implies that the evolution of a Gaussian state
can be obtained, if the time dependence of these parameters is known. In this
work, we review the differential equations to which the covariance
matrix and the mean values satisfy~\cite{vdod1,vdod2}; employing these results
we can define differential equations for the density matrix parameters of a
general multimode state satisfying these equations, and then use the equations
to discuss some physical characteristics of the unitary and nonunitary
evolutions of Gaussian states.

This paper is organized as follows.

In section~2, the evolution of non-pure Gaussian states for a one-dimensional
quadratic Hamiltonian is presented. To obtain this evolution, we make use of
the derivatives of the covariance matrix, the mean values, and the parameters
of the density operator; also we define and obtain invariant states for this
system. The generalization of these results to the case of a multidimensional
quadratic system is explored in section~3. As examples of the application of
the general results  to the nonunitary evolution of the subsystems of a
two-mode state, as well as the definition of invariant and quasi-invariant
states is done in section~4. Also in section~5, we obtain new invariant states
for the frequency converter and quasi-invariant states for the parametric
amplifier. The detection of this invariant states using the quantum
tomographic representation of the states is discussed for single-mode Gaussian
states in section 5 and for the bipartite system in section 6; in these
sections, the correspondence between the time-independent states and thermal
density matrices is mentioned. Finally, we give our conclusions.

\section{One-Dimensional Quantum Quadratic Hamiltonian and Its Linear Invariant Operators}

In this section, we analyze some properties of the one-dimensional
quadratic Hamiltonian. Particularly, we are interested in the
invariant operators, which in the quadratic case happen to be linear
in the quadrature operators $\hat{p}$ and $\hat{q}$.

The most general (in a unit system where $\hbar=m=1$), one-dimensional quantum quadratic Hamiltonian can
be obtained in terms of the quadrature operators $\hat{p}$ and
$\hat{q}$ as follows:
\begin{equation}
\hat{H}= (\hat{p},\hat{q})
\left(
\begin{array}{cc}
\omega_1(t) & \omega_2(t) \\
\omega_2(t) & \omega_3(t)
\end{array}
\right)
\left(
\begin{array}{cc}
\hat{p} \\
\hat{q}
\end{array}
\right)+(\hat{p},\hat{q})
\left(
\begin{array}{cc}
\delta_1 \\
\delta_2
\end{array}
\right)
\, ,
\label{ham}
\end{equation}
where the parameters $\omega_1(t)$, $\omega_2(t)$, $\omega_3(t)$, and
$\delta_{1,2}$ are real functions of time. The dynamics associated
to this Hamiltonian can be solved by different methods. One of them
is the method of time-dependent invariants (integrals of
motion)~\cite{inv1,inv2}. These invariants are quantum operators
$\hat{R}(t)$, whose total time derivative is equal to zero
$\dfrac{d\hat{R}(t)}{dt}=0$. In the quadratic case, it is known that
there exist invariants linearly depending on the quadrature
operators $\hat{p}$ and $\hat{q}$, i.e., $\hat{R}(t)=\lambda_1(t)
\hat{p}+\lambda_2 (t)
\hat{q}+\lambda_3(t)$.

By substituting this expression into the von Neumann equation, which
determines the dynamics of $\hat{R}$, i.e., $
\dfrac{d\hat{R}(t)}{dt}=\dfrac{i}{\hbar}[\hat{H}(t),\hat{R}(t)]+\dfrac{\partial
\hat{R}(t)}{\partial t}=0$, one can show that $\hat{R}(t)$ is an invariant
operator, if the following differential equations are satisfied:
\begin{eqnarray}
\dot{\lambda}_1=2(\omega_{2} \lambda_1- \omega_1 \lambda_2) \, ,
\qquad \dot{\lambda}_2=2(- \omega_{2} \lambda_2+\omega_3 \lambda_1) \,
,\qquad \dot{\lambda}_3=\delta_2 \lambda_1-\delta_1 \lambda_2 \, .
\label{quan}
\end{eqnarray}
We point out that parameters $\lambda_{1,2}$ satisfy the classical Hamilton equations
\begin{eqnarray}
\dot{p}=-2(\omega_{2} p+ \omega_3 q)-\delta_2 \, , \qquad
\dot{q}=2(\omega_{2} q+\omega_1 p) +\delta_1\, .
\label{clas}
\end{eqnarray}
with $\delta_1=\delta_2=0$. To show this, one can see that the differential
equations for $\lambda_{1,2}$ correspond to the classical equations with the
substitution $\lambda_1 \rightarrow q$ and $\lambda_2 \rightarrow -p$; in
other words, they correspond to the time inversion of the classical equations.
In the case of the differential equation for $\lambda_3$, one can show,
in view of the Hamilton equations, that it corresponds to
the classical Lagrangian (with $\delta_{1,2} \neq 0$) plus the
time variation of the function $pq$ of the system, that is,
\begin{equation}
\dot{\lambda}_3=-2 \mathcal{L}+\dot{L} \, ,
\end{equation}
where $\mathcal{L}=p \dot{q}-H$ and $L=pq$. From these
identifications, one can conclude that the classical dynamics given
by the Hamilton equation or the equation of motion can lead to the
solution of quantum dynamics given by the Hamiltonian operator of
equation~(\ref{quan}). For example, one can derive the propagator of
the system $G(x,x',t)=\langle x \vert \hat{U}(t) \vert x'
\rangle$ using each one of the solutions of the classical
problem~\cite{Dodon-}.

\subsection{Dynamics of Non-Pure States}
Here, we demonstrate that the dynamics of a generic Gaussian state, which
may be not pure, can be given by solving differential equations for the
covariance matrix or the density matrix parameters. We show that these
differential equations imply the invariance of the determinant of the
covariance matrix, when the time evolution is unitary.

The propagator of the system can be obtained using the time-dependent
invariants resulting of the solution to equation~(\ref{quan}) for two sets of
initial conditions: $\lambda_1(0)=1$, $\lambda_2(0)=0$, $\lambda_3(0)=0$ and
$\lambda_1(0)=0$, $\lambda_2(0)=1$, $\lambda_3(0)=0$. These two sets define
two different invariants called $\hat{P}$ and $\hat{Q}$, respectively, which
can be written as
\begin{equation}
\left(
\begin{array}{cc}
\hat{P} \\
\hat{Q}
\end{array}
\right)= \boldsymbol{\Lambda}
\left(
\begin{array}{cc}
\hat{p} \\
\hat{q}
\end{array}
\right)+
\left(
\begin{array}{cc}
\lambda_3 \\
\lambda_6
\end{array}
\right)\, , \quad {\rm with} \quad
\boldsymbol{\Lambda}=\left(\begin{array}{cc}
\lambda_1 & \lambda_2 \\
\lambda_4 & \lambda_5
\end{array}\right) \, ,
\label{invg}
\end{equation}
where $\lambda_{4,5,6}$ satisfy the same differential equations that
$\lambda_{1,2,3}$, respectively, with the different sets of initial
conditions mentioned above. The operators $\hat{P}$ and $\hat{Q}$
fulfill the commutation relation $[\hat{Q},\hat{P}]=i$, implying the
relation $\lambda_1 \lambda_5-\lambda_2 \lambda_4=1$ and the fact
that the matrix $\boldsymbol{\Lambda}$ is symplectic.

Also, they can be related to operators $\hat{p}$ and $\hat{q}$
through the evolution operator as follows:
\[
\hat{P}=\hat{U} \hat{p}\, \hat{U}^\dagger \, , \qquad \hat{Q}=\hat{U} \hat{q}\, \hat{U}^\dagger \,
;
\]
it is not difficult to show~\cite{inv1,inv2} that these expressions
can be used for obtaining the propagator of the system
$G(x,x',t)=\langle x
\vert \hat{U}(t) \vert x' \rangle$, which reads
\begin{eqnarray}
G(x,x',t) &=&\frac{1}{\sqrt{-2\pi i \lambda_4}}\,
\exp\left\{-\frac{i}{2
\lambda_4}\Huge[\lambda_5 x^2- 2 x x'+\lambda_1 x^{\prime2} +2x
\lambda_6+2 x'(\lambda_3 \lambda_4-\lambda_1 \lambda_6)
\right.\nonumber\\
&&
\left.~\qquad\qquad\qquad
+\lambda_1
\lambda_6^2 -2 \lambda_4 \int_0^t \dot{\lambda}_3 \lambda_6
d\tau\LARGE{]} \right\} ,
\end{eqnarray}
and which we immediately identify as a Gaussian function.

In view of this propagator, the dynamics of any initial state in the position
representation can be found by the integration of the propagator and the wave
function of the initial state. In this work, we will suppose the case of the
initial state given by a generic mixed Gaussian system with density matrix in
the position representation $\rho(x,x',t=0)=\langle x \vert \hat{\rho} \vert
x' \rangle$ equal to
\begin{equation}
\rho(x,x',t=0)=N \exp\left\{-a_1 x^2+a_{12} x x'-a_1^* x^{\prime2}+b_1 x+b_1^* x' \right\} \, ,
\label{gaus0}
\end{equation}
with complex parameters $a_1=a_{1R}+ia_{1I}$ and $b_1=b_{1R}+i
b_{1I}$ and the real parameter $a_{12}\in \mathbb{R}$, which also
satisfy the integrability conditions $a_{1R} > a_{12}/2\geq 0$ and
has a normalization constant $N$ expressed as
\[
N=\left( \frac{a_1+a_1^*-a_{12}}{\pi}\right)^{1/2}
\exp\left\{-\frac{(b_1+b_1^*)^2}{4(a_1+a_1^*-a_{12})}\right\}\, .
\]
As discussed above, the Gaussian states can be fully identified by
their covariance matrix and mean values of the quadrature
components. In the case of state (\ref{gaus0}), the mean values of the operators $\hat{p}$ and
$\hat{q}$ are
\begin{equation}
\langle \hat{p} \rangle(0)=b_{1I}-\frac{2a_{1I} b_{1R}}{2a_{1R}-a_{12}}
\, , \qquad \langle \hat{q} \rangle(0)=\frac{b_{1R}}{2a_{1R}-a_{12}} \, ,
\end{equation}
and the initial covariance matrix of the system reads
\begin{equation}
\boldsymbol{\sigma}(0)=
\left(\begin{array}{cc}
\sigma_{pp} & \sigma_{pq} \\
\sigma_{pq} & \sigma_{qq}
\end{array}\right)
=\frac{1}{2\,(2a_{1R}-a_{12})}\left(\begin{array}{cc}
4 \vert a_1 \vert^2-a_{12}^2 & -2 a_{1I} \\
-2 a_{1I}  & 1
\end{array}\right) .
\label{covv}
\end{equation}
Here, the covariance between arbitrary operators $\hat{x}$ and $\hat{y}$ is
given in terms of the expectation value of the anticommutator, i.e.,
$\sigma_{xy}=\dfrac{1}{2}\,{\rm
Tr}\,(\hat{\rho}(\hat{x}\hat{y}+\hat{y}\hat{x}))-{\rm
Tr}\,(\hat{\rho}\hat{x})\,{\rm Tr}\,(\hat{\rho}\hat{y})$.

All properties of the Gaussian state can be obtained by the use of
the covariance matrix and the mean values of the state. For example,
the purity of the Gaussian state can be obtained by the determinant
of its covariance matrix, this is,
\begin{equation}
{\rm Tr}\, \hat{\rho}^2= \frac{1}{2\sqrt{\det\boldsymbol{\sigma}}} \, .
\end{equation}

The unitary dynamics of the initial state of equation~(\ref{gaus0})
can be obtained by the integration of the propagators multiplied by
the mixed-state density matrix
\[
\rho(x,x',t)=\int_{-\infty}^\infty \int_{-\infty}^\infty dx_1\,
dx_2\,
\, G^*(x_1,x,t) \rho(x_1,x_2,0)\, G(x_2,x',t) \, ;
\]
as a result, it provides a Gaussian state with the same purity as the original
state (${\rm Tr}\, \hat{\rho}^2(t)={\rm Tr}\, \hat{\rho}^2(0)$), since unitary
transforms do not change purity, which then can infer the determinant
invariance of the covariance matrix $\det \boldsymbol{\sigma}(0)=\det
\boldsymbol{\sigma}(t)$.

In a similar way, we can write the final state in an analogous way as the initial one, this is
\begin{equation}
\rho(x,x',t)=N \exp\left\{-a_1(t) x^2+a_{12}(t) x x'-a_1^*(t)
x^{\prime2}+b(t) x+b(t)^* x' \right\} \, ,
\label{stat}
\end{equation}
where the Gaussian parameters are written in terms of the symplectic
matrix associated with the invariants of equation~(\ref{invg});
thus, we arrive at the following expressions:
\begin{eqnarray}
a_1(t)&=&\frac{1}{2 \lambda_4} \left( \frac{2 a_1^* \lambda_4
-i \lambda_1}{\lambda_1^2-2 i(a_1-a_1^*)\lambda_1 \lambda_4
+d \lambda_4^2}+i\lambda_5\right) \, , \nonumber \\
a_{12}(t)&=&\frac{a_{12}}{\lambda_1^2-2 i(a_1-a_1^*)\lambda_1
\lambda_4+d \lambda_4^2} \, , \\
b(t)&=&\frac{\lambda_1(b-i \lambda_3+(a_{12}-2 a_1)\lambda_6)
+\lambda_4((2a_1^*-a_{12})\lambda_3+i(2a_1^* b+a_{12}b^* -\delta
\lambda_6)) }{\lambda_1^2-2 i(a_1-a_1^*)\lambda_1 \lambda_4
+d\lambda_4^2},
 \nonumber
\end{eqnarray}
with $d=4 \vert a_1 \vert^2-a_{12}^2$.

It is possible to obtain the differential equations, which these parameters
must satisfy. This is done by taking the time derivative of the parameters
and, in view of equation~(\ref{quan}); after some algebra, we obtain
\begin{eqnarray}
\dot{a}_1(t)&=& i (a_{12}^2 (t) - 4 a_1^2(t)) \omega_1(t) - 4 a_1(t) \omega_{2}(t) + i \omega_3(t),  \nonumber \\
\dot{a}_{12}(t)&=&4 a_{12} (t)( i (a_1^*(t) - a_1(t)) \omega_1(t) -  \omega_{2}(t)), \label{nonli}  \\
\dot{b} (t)&=&(2 a_1(t)-a_{12}(t) ) \delta_1(t) - i \delta_2(t) - 2 i
a_{12}(t) b^*(t) \omega_1(t) -2
 b(t) ( \omega_{2}(t)+2 i a_1(t) \omega_1(t)) \, ,\nonumber
\end{eqnarray}
and their corresponding complex conjugates. It is worth mentioning that these
equations can be corroborated by the use of the von Neumann equation for
$\rho(x,x',t)$ given in equation~(\ref{stat}), namely, $~i\,\dfrac{d\rho
(x,x',t)}{dt}=\langle x \vert [\hat{H}, \hat{\rho}]\vert x'\rangle$. On the
other hand, it is known that the  covariance matrix of the system can be
obtained, in view of the quantum solutions of equation~(\ref{quan}); as we
have pointed out, this corresponds to the classical solutions~(\ref{clas})
with $\delta_1=\delta_2=0$, which can be also written in terms of the
symplectic transformation $\boldsymbol{\Lambda}$ of equation~(\ref{invg}),
i.e., $\boldsymbol{\sigma} (t)=\boldsymbol{\Lambda}^{-1} \boldsymbol{\sigma}
(0)\tilde{\boldsymbol{\Lambda}}^{-1}$. Then the covariance matrix evolution
$\boldsymbol{\sigma}(t)$ can be obtained as
\begin{equation}
\boldsymbol{\sigma}(t)=
\left(\begin{array}{cc}
\sigma_{pp} (t) & \sigma_{pq}(t) \\
\sigma_{pq}(t) & \sigma_{qq} (t)
\end{array}\right)
=
\left(\begin{array}{cc}
\lambda_5 & -\lambda_2 \\
-\lambda_4 & \lambda_1
\end{array}\right)
\left(\begin{array}{cc}
\sigma_{pp} (0) & \sigma_{pq}(0) \\
\sigma_{pq}(0) & \sigma_{qq} (0)
\end{array}\right)
\left(\begin{array}{cc}
\lambda_5 & -\lambda_4 \\
-\lambda_2 & \lambda_1
\end{array}\right)\, .
\label{cov_t}
\end{equation}

After differentiating each covariance $\dot{\sigma}_{pp} (t)$,
$\dot{\sigma}_{qq} (t)$, and $\dot{\sigma}_{pq} (t)$, by using equations~(\ref{quan}), the inverse expression of
equation~(\ref{cov_t}), the purity conservation condition
$\sigma_{pp}(0)\sigma_{qq}(0)-\sigma_{pq}^2
(0)=\sigma_{pp}(t)\sigma_{qq}(t)-\sigma_{pq}^2 (t)$, and the
condition $\lambda_1 \lambda_5-\lambda_2 \lambda_4=1$, we arrive at
the following differential equations for the covariances:
\begin{eqnarray}
&&\dot{\sigma}_{pp}(t)=-4\,(\omega_{2}(t)\sigma_{pp}(t)+\omega_3(t)
\sigma_{pq}(t)) \, ,\nonumber\\
&&\dot{\sigma}_{qq}(t)=4\,(\omega_{2}(t)\sigma_{qq}(t)+\omega_1(t) \sigma_{pq}(t)), \\
&&\dot{\sigma}_{pq}(t)=2\,(\omega_{1}(t)\sigma_{pp}(t)-\omega_3(t)
\sigma_{qq}(t)) \, .\nonumber
\label{diff1}
\end{eqnarray}
One can also check that these differential equations imply that the
derivative of the determinant of $\boldsymbol{\sigma}(t)$ is equal
to zero, i.e., $\dfrac{d}{dt}\left[\sigma_pp (t)
\sigma_{qq}(t)-\sigma_{pq}^2\right]=0$, which also implies that the
purity of state~(\ref{stat}) is a time invariant. It is noteworthy
that the time-derivative expressions for the covariance matrix can
be expressed as follows:
\begin{equation}
\dot{\boldsymbol{\sigma}}(t)=2\left[\boldsymbol{\sigma}(t) \mathbf{B}(t)
\boldsymbol{\Sigma}- \boldsymbol{\Sigma} \mathbf{B}(t) \boldsymbol{\sigma}(t)\right] \, ,
\label{covg}
\end{equation}
where the matrix $\mathbf{B}(t)$ contains the Hamiltonian
coefficients, while $\boldsymbol{\Sigma}$ is a symplectic matrix,
i.e.,
\[
\mathbf{B}(t)=\left(
\begin{array}{cc}
\omega_1 (t) & \omega_2 (t)\\
\omega_2 (t) & \omega_3 (t)
\end{array}
\right) \, , \qquad
\boldsymbol{\Sigma}=
\left(\begin{array}{cc}
0 & 1 \\
-1 & 0
\end{array}
\right) \, .
\]
On the other hand, one can also check, using the inverse expression
of equation~(\ref{invg}), that the mean values of $\hat{p}$ and
$\hat{q}$ follow the classical equation~(\ref{clas}), i.e.,
\begin{equation}
\frac{d}{dt}\left( \begin{array}{cc}
\langle \hat{p} \rangle \\
\langle \hat{q} \rangle
\end{array}\right)=
2\left(\begin{array}{cc}
-\omega_2 & -\omega_3 \\
\omega_1 & \omega_2
\end{array}\right)
\left( \begin{array}{cc}
\langle \hat{p} \rangle \\
\langle \hat{q} \rangle
\end{array}\right)+
\left(\begin{array}{cc}
-\delta_1 \\
\delta_2
\end{array}
\right) \, .
\label{meang}
\end{equation}

All information regarding the evolution of the Gaussian state can
then be obtained by solving the differential equations~(\ref{covg})
and (\ref{meang}). As an example, we can consider the evolution of a
Gaussian state with the initial covariance matrix
$\boldsymbol{\sigma}(0)$ and mean values $\langle \hat{p} \rangle
(0)$ and $\langle \hat{q} \rangle (0)$.

\begin{figure}
\begin{subfigure}[]{
\includegraphics[scale=0.4]{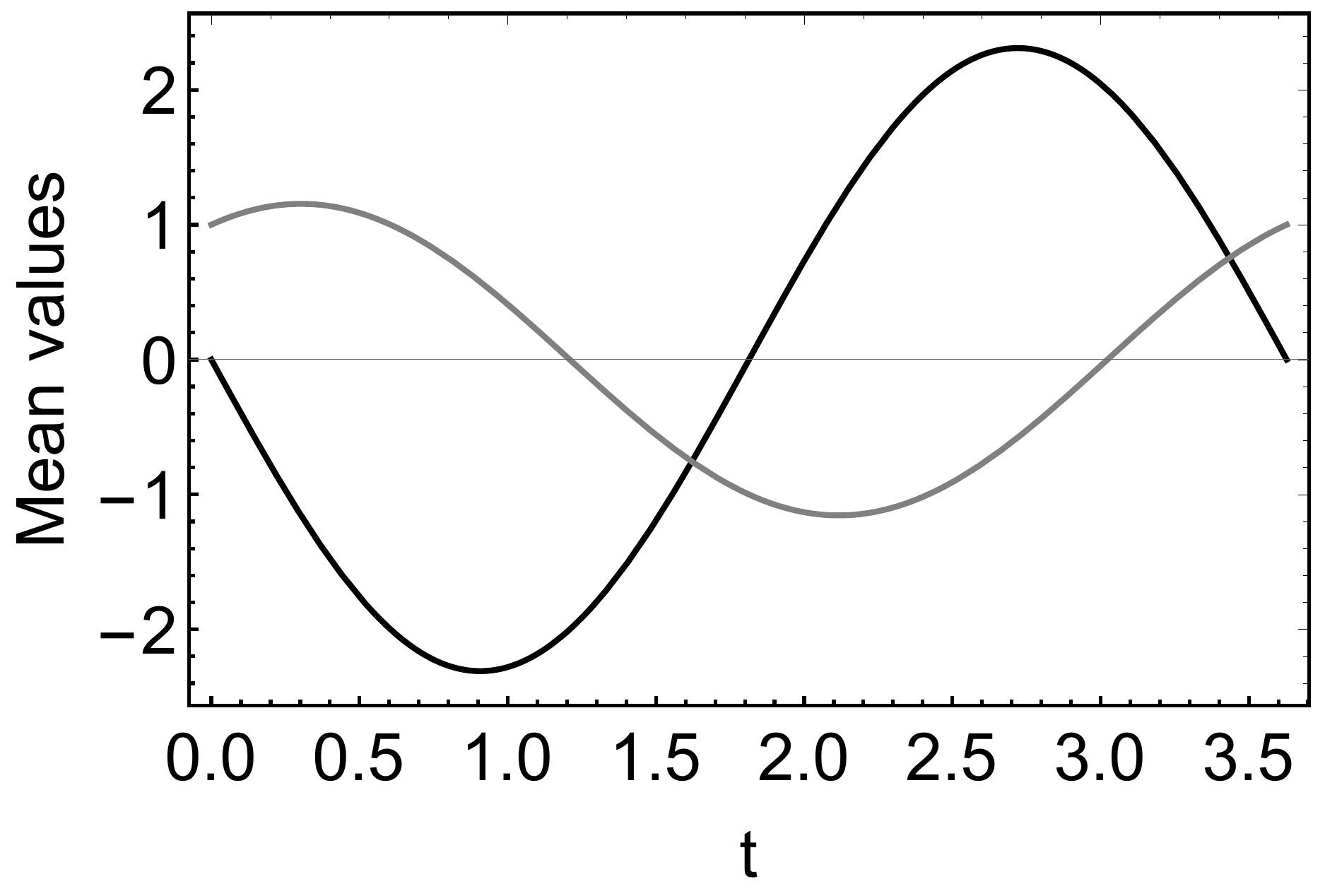}}
\end{subfigure}
\begin{subfigure}[]{
\includegraphics[scale=0.4]{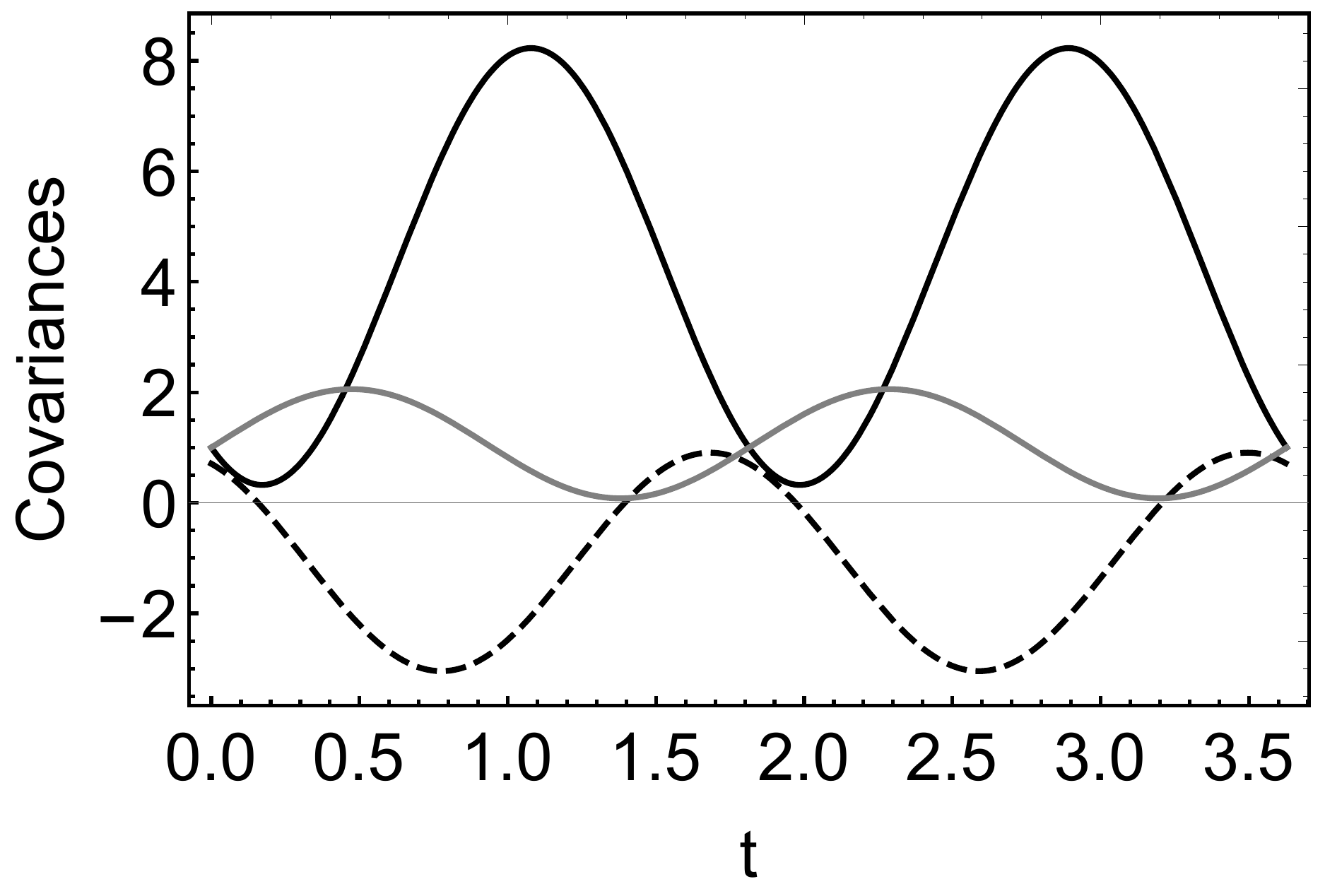}}
\end{subfigure}
\caption{(a)~Mean values $\langle \hat{p} \rangle (t)$~(black) and
$\langle \hat{q} \rangle (t)$~(gray) for the dynamics of
Hamiltonian~(\ref{ham_ex1d}) and the state with initial conditions
$\langle \hat{p} \rangle (0)=0$ and $\langle \hat{q} \rangle (0)=1$.
(b)~Covariances $\sigma_{pp}(t)$~(black), $\sigma_{qq}(t)$~(gray),
and $\sigma_{pq}(t)$~(dashed) for the initial state with
$\sigma_{pp}(0)=\sigma_{qq}=1$ and $\sigma_{pq}(0)=1/\sqrt{2}$. In
both cases, we took frequencies $\omega=2$ and $\nu=1$.\label{fig1}}
\end{figure}

\subsubsection{Example}
As an example, we consider the following
Hamiltonian:
\begin{equation}
\hat{H}=\frac{1}{2}(\hat{p}^2+\omega^2 \hat{q}^2)+\frac{\nu}{2} (\hat{p}
\hat{q}+\hat{q} \hat{p}) \, .\label{ham_ex1d}
\end{equation}
In view of equations~(\ref{covg}) and (\ref{meang}), the matrix $\mathbf{B}(t)$ can be identified as
\[
\mathbf{B}=\frac{1}{2}\left(\begin{array}{cc}
1 & \nu \\
\nu & \omega^2
\end{array}\right) \, ,
\]
and one can show that, in the case of constant frequencies
($\omega$, $\nu$), the evolution is determined by the same differential
equations for all the covariances
\[
\dddot{\sigma}_{pp}=-4( \omega^2-\nu^2) \dot{\sigma}_{pp}\, , \quad \dddot{\sigma}_{qq}=-4( \omega^2-\nu^2) \dot{\sigma}_{qq}\, , \quad \dddot{\sigma}_{pq}=-4( \omega^2-\nu^2) \dot{\sigma}_{pq} \, ,
\]
which at $\omega^2>\nu^2$ describes an oscillating motion with parameter
$f^2=4(\omega^2-\nu^2)$. The solution to these equations, which satisfy the
initial conditions for the derivatives and second derivatives at time $t=0$
implied by equation~(\ref{diff1}), are
\begin{eqnarray}
& \sigma_{pp}(t)=-\frac{2 \left(f \sin (f t) \left(w^2 {z_0}+\nu  {x_0}\right)+\cos (f t)
\left(\omega^4 {y_0}-\omega^2 ({x_0}-2 \nu  {z_0})+2 \nu ^2 {x_0}\right)
-\omega^2 \left(\omega^2 {y_0}+{x_0}+2 \nu  {z_0}\right)\right)}{f^2} \, , \nonumber \\
&\sigma_{qq}(t)=\frac{2 \left(-\cos (f t) \left(\omega^2 (-{y_0})+{x_0}
+2 \nu  (\nu  {y_0}+{z_0})\right)+f \sin (f t) (\nu  {y_0}+{z_0})+\omega^2 {y_0}+{x_0}
+2 \nu  {z_0}\right)}{f^2} \, , \label{sol_ex1d} \\
 &\sigma_{pq}(t)=\frac{2 \cos (f t) \left(\omega^2 (\nu  {y_0}+2 {z_0})
 +\nu  {x_0}\right)+f \sin (f t) \left({x_0}-\omega^2 {y_0}\right)
 -2 \nu  \left(\omega^2 {y_0}+{x_0}+2 \nu  {z_0}\right)}{f^2} \, ,\nonumber
\end{eqnarray}
with $x_0=\sigma_{pp}(0)$, $y_0=\sigma_{qq}(0)$, and $z_0=\sigma_{pq}(0)$. It
is worth mentioning that these expressions contain information on the time
invariance of the determinant of $\boldsymbol{\sigma}(t)$, since it can be
checked that
$\sigma_{pp}(t)\sigma_{qq}(t)-\sigma_{pq}^2(t)=\sigma_{pp}(0)\sigma_{qq}(0)-\sigma_{pq}^2(0)$.

The solution for the classical equations of motion for the mean
values reads
\begin{eqnarray}
\langle \hat{p} \rangle (t)&=&\langle \hat{p} \rangle (0)
\cos \left({f t}/{2}\right)-({2}/{f}) \left[\omega^2 \langle \hat{q} \rangle (0)
+\nu \langle \hat{p} \rangle (0)\right] \sin\left({f t}/{2}\right)\, , \nonumber
\\[-2mm]
\label{mean_ex1d}\\[-2mm]
\langle \hat{q} \rangle (t)&=&\langle \hat{q} \rangle (0)
\cos \left({f t}/{2}\right)+({2}/{f}) \left[\nu \langle \hat{q} \rangle (0)+
\langle \hat{p} \rangle (0)\right] \sin\left({f t}/{2}\right)\, .
\nonumber
\end{eqnarray}
With these solutions, one can characterize the state behavior. In
Fig.~\ref{fig1}, we show the evolution of the mean values and covariance
matrix given by equations~(\ref{sol_ex1d}) and~(\ref{mean_ex1d}) for the
Hamiltonian~(\ref{ham_ex1d}). Here, we observe the oscillating behavior of the
system. We would like to remark that, using this solutions for the
covariances and the correspondence between the density matrix parameters and
the covariances:
\begin{equation}
\boldsymbol{\sigma}(t)=\frac{1}{2(a_1(t)+a_1^*(t)-a_{12}(t))}\left(
\begin{array}{cc}
4 a_1(t) a_1^*(t)-a_{12}^2(t) & i(a_1(t)-a_1^*(t)) \\
i(a_1(t)-a_1^*(t)) & 1
\end{array}
\right),
\end{equation}
one can also obtain the solution for the nonlinear equations~(\ref{nonli}).

\subsection{Invariant States}
After obtaining the differential equations determining the evolution of the
covariance matrix~(\ref{covg}), one can put the question: Do invariant states
exist under the evolution of the quadratic Hamiltonian? To answer this
question, we should examine the properties of equation~(\ref{covg}). If we
assume the condition $\dot{\boldsymbol{\sigma}}(t)=0$, then one needs to
obtain all the covariance matrices, which satisfy the condition
$\boldsymbol{\sigma}(t) \mathbf{B}(t) \boldsymbol{\Sigma}- \boldsymbol{\Sigma}
\mathbf{B}(t) \boldsymbol{\sigma}(t)=0$. By taking the vector
$\tilde{\mathbf{v}}=(\sigma_{pp}, \sigma_{pq}, \sigma_{qq})$, the equations
for $\dot{\boldsymbol{\sigma}}(t)=0$ can be written as follows:
\begin{equation}
\mathbf{Mv}=0, \quad {\rm with} \quad \mathbf{M}=\left( \begin{array}{ccc}
-4 \omega_2 & -4 \omega_3 & 0 \\
2 \omega_1 & 0 & -2 \omega_3 \\
0 & 4 \omega_1 & 4 \omega_2
\end{array}
\right) \, .
\label{inv1d}
\end{equation}
As matrix $\mathbf{M}$ has rank $\mathcal{R}=2$, one can conclude
that there is one nontrivial vector satisfying
equation~(\ref{inv1d}). Exploring the null-space of $\mathbf{M}$ one
can check that the vector
\[
\tilde{\mathbf{v}}=C\left(\frac{\omega_3}{\omega_1}, -\frac{\omega_2}{\omega_1}, 1\right) \, ,
\]
with $C$ being a constant, is the solution to equation~(\ref{inv1d}). We conclude that all
the states with a covariance matrix, given by
\begin{equation}
\boldsymbol{\sigma}(0)=C\left(\begin{array}{cc}
\dfrac{\omega_3}{\omega_1} & -\dfrac{\omega_2}{\omega_1} \\
-\dfrac{\omega_2}{\omega_1} & 1
\end{array}\right),
\end{equation}
have invariant covariance matrix. Using the inverse expressions of
equations~(\ref{covv}), one can obtain an explicit form of the
covariance invariant density matrix function. The parameters of the
density operator~(\ref{stat}) read
\begin{equation}
a_1=\frac{4 C^2 \omega_1  \omega_3+(\omega_1+2 i C \omega_2)^2}{8 C \omega_1^2} \, , \quad a_{12}= \frac{4 S-1}{4C} \, ,
\label{inv_a}
\end{equation}
with $S=C^2(\omega_3/\omega_1-\omega_2^2/\omega_1^2)$ being the determinant of
the invariant covariance matrix. In the case of Hamiltonian~(\ref{ham_ex1d}),
we have $S=C^2(\omega^2-\nu^2)$, $\omega_3/\omega_1=\omega^2$, and
$\omega_2/\omega_1=\nu$, which lead us to realize that any state with
parameters, as in equation~(\ref{inv_a}) with $S>1/4$, are bonafide quantum
states which are covariance invariant. For $C=1$, $\omega=2$, and $\nu=1$, we
have that the states with
\begin{equation}
a_1=\frac{13}{8}+\frac{i}{2}, \quad a_{12}=\frac{11}{4} \label{para1d}
\end{equation}
are covariance invariant.

The states, which satisfy $\dot{\boldsymbol{\sigma}}=0$ or equivalently have
parameters according to~(\ref{inv_a}), are the states which do not change its
shape on the phase space $q,p$; also their mean values move following the
classical equations of motion. If we assume these type of states with mean
values $\langle \hat{p}\rangle (0)=\langle \hat{q} \rangle (0)=0$, the
resulting states will be invariant, i.e., they will not change any of their
properties over time (for $\delta_1=\delta_2=0$). An example of such states
for the Hamiltonian~(\ref{ham_ex1d}) are the ones in equation~(\ref{stat}),
with parameters given by~(\ref{para1d}) and $b(t)=b^*(t)=0$. We would like to
express that, in the case of an invariant system with vanishing mean values,
the initial energy will be different from zero as the initial covariances are
also different from zero.

This parametric formalism for the evolution of Gaussian states and
the definition of invariant states can be generalized to any
multidimensional quadratic system as seen in the following section.


\section{Multidimensional Quadratic System}
In this section, we review the equations determining the evolution of the
covariance matrix and mean values $\langle p_j \rangle$ and $\langle q_j
\rangle$ for an arbitrary system under the evolution of a quadratic
Hamiltonian; also we mention the connection and dynamics of the continuous
density matrix parameters. To obtain these properties, we use, as in the
one-dimensional case, the invariant operators defined in~\cite{inv1,inv2}.

In the case of an $N$-dimensional quadratic system, the time evolution is
characterized by the Hamiltonian
\begin{equation}
\hat{H}= \tilde{\mathbf{r}}\mathbf{B}(t) \mathbf{r}
+\tilde{\boldsymbol{\Delta}}(t) \mathbf{r}\, ,
\label{ham-mult}
\end{equation}
where the tilde corresponds to the transposition operation, the
vector $\tilde{\mathbf{r}}=\left(\hat{p}_1, \hat{q}_1, \hat{p}_2,
\hat{q}_2, \ldots, \hat{p}_N, \hat{q}_N \right)$ corresponds to the
vector of quadrature operators. The time dependence of this
Hamiltonian is contained in the matrices
\begin{equation}
\mathbf{B}(t)=\left( \begin{array}{cccc}
\omega_{1,1}(t) & \omega_{1,2}(t) & \cdots & \omega_{1,2N}(t) \\
\omega_{1,2}(t) & \omega_{2,2}(t) & \cdots & \omega_{2,2N}(t) \\
\vdots & \vdots & \ddots & \vdots \\
\omega_{1,2N}(t) & \omega_{2,2N}(t) & \cdots & \omega_{2N,2N} (t)
\end{array}\right) , \qquad
\boldsymbol{\Delta}(t)=\left( \begin{array}{cccc}
\delta_1 (t) \\
\delta_2(t)  \\
\vdots  \\
\delta_{2N}(t)
\end{array}\right) \, ,
\end{equation}
where $\mathbf{B}(t)$ is a real and symmetric matrix and
$\boldsymbol{\Delta}(t)$ is a real vector. As in the one-dimensional case,
there exist $2N$ linear time-dependent operators $\hat{R}_j$
($j=1,\ldots,2N$), whose time derivatives are equal to zero
$\left(\dfrac{d\hat{R}_j}{dt}=0.\right)$. These operators can be arranged on a
vector as follows:
\begin{equation}
\boldsymbol{\mathbf{R}}=\boldsymbol{\Lambda}(t) \mathbf{r}+ \boldsymbol{\Gamma}(t) \, ,
\label{inv4}
\end{equation}
with the matrix $\boldsymbol{\Lambda}(t)$ and the vector
$\tilde{\boldsymbol{\Gamma}}(t)=\left(\gamma_1 (t), \gamma_2 (t), \ldots,
\gamma_{2N} (t) \right)$. By taking the time-derivative of the operator
$\boldsymbol{\mathbf{R}}_j=\hat{R}_j$ and equating it to zero
$\left(\dfrac{d\hat{R}_j}{dt}=0.\right)$, one can demonstrate that
$\boldsymbol{\Lambda}(t)$ and $\boldsymbol{\Gamma}(t)$ must satisfy the
following differential equations:
\begin{equation}
\dot{\boldsymbol{\Lambda}}(t)=2i\boldsymbol{\Lambda}(t) \mathbf{D}
\mathbf{B}(t), \quad \dot{\boldsymbol{\Gamma}}(t)=i \boldsymbol{\Lambda}(t)
\mathbf{D} \boldsymbol{\Delta}(t), \quad {\rm with}\quad \mathbf{D}_{j,k}=
[ \mathbf{r}_j, \mathbf{r}_k ] \, .
\label{lambdad}
\end{equation}
The solution to these differential equations with the initial conditions
$\hat{R}_j(0)=\mathbf{r}_j$ provides, as a result, the invariant operators
$\tilde{\boldsymbol{R}}=\left(\hat{P}_1,\hat{Q}_1,\hat{P}_2,\hat{Q}_2,
\ldots,\hat{P}_N,\hat{Q}_N\right)$, satisfying the standard commutation rules
for the operators at zero time: $\mathbf{r}$, i.e., $[ \mathbf{R}_j,
\mathbf{R}_k]=[\mathbf{r}_j, \mathbf{r}_k]=\mathbf{D}_{j,k}$. This property
leads us to the conclusion that the matrix $\boldsymbol{\Lambda}(t)$ must be
symplectic and satisfies the equation
\begin{equation}
\boldsymbol{\Lambda}(t) \mathbf{D} \tilde{\boldsymbol{\Lambda}}(t)=
\mathbf{D},
\end{equation}
with $\mathbf{D}_{j,k}=[\mathbf{r}_j,\mathbf{r}_k]$; this relation
can be then used to obtain the inverse of $\boldsymbol{\Lambda}(t)$,
which results in the expression
\begin{equation}
\boldsymbol{\Lambda}^{-1}(t)= \mathbf{D} \tilde{\boldsymbol{\Lambda}}(t)
\mathbf{D} \, .
\label{inversel}
\end{equation}

The other important property of these invariant operators is that
they correspond to the inverse evolution of the original operators, in other words,
\[
\mathbf{R}_j=\hat{U}(t) \mathbf{r}_j \hat{U}^\dagger (t),
\]
which in the most cases can be obtained from the Heisenberg picture
operators by assuming a time reversal operation. This property
implies that the entries of $\dot{\boldsymbol{\Lambda}}(t)$ in
equation~(\ref{lambdad}) satisfy the classical Hamilton equations
after the time reversal operation, that is, after the change $p_i \rightarrow -p_i$ in the classical Hamilton equations.

By the use of these invariant operators, one can obtain the time dependence of
the mean values of the operators in $\mathbf{r}$ ($\langle \mathbf{r}\rangle
(t)$) and their covariances $\boldsymbol{\sigma}_{j,k}=\dfrac{1}{2}\langle \{
\mathbf{r}_j,\mathbf{r}_k \}\rangle -\langle \mathbf{r}_j \rangle \langle
\mathbf{r}_k \rangle$. From the inverse of equation~(\ref{inv4}), one can
demonstrate that
\[
\langle \mathbf{r} \rangle(t)= \boldsymbol{\Lambda}^{-1}(t)
(\langle \mathbf{R} \rangle -\boldsymbol{\Gamma}(t)) \, ,
\]
as the invariant operators in $\mathbf{R}$ have a time derivative equal to
zero, and they are equal to the standard operators $\mathbf{r}$ at zero time,
then one can conclude that
\begin{equation}
\langle \mathbf{r} \rangle(t)= \boldsymbol{\Lambda}^{-1}(t) (\langle
\mathbf{r} \rangle (0) -\boldsymbol{\Gamma}(t)) \, .\label{meant}
\end{equation}
From an analogous argument, one can see that the covariance matrix
reads
\begin{equation}
\boldsymbol{\sigma}(t)=\boldsymbol{\Lambda}^{-1}(t)
\boldsymbol{\sigma}(0) \tilde{\boldsymbol{\Lambda}}^{-1}(t) \, .
\label{sigmat}
\end{equation}
Then to obtain the expression for the time-derivative of the mean
values $\langle \mathbf{r} \rangle(t)$ and the covariance matrix
$\boldsymbol{\sigma}(t)$, we make use of equations~(\ref{lambdad}),
(\ref{inversel}), (\ref{meant}), and (\ref{sigmat}) and arrive at
the expressions
\begin{equation}
\frac{d}{dt} \langle \mathbf{r} \rangle(t)= -i \mathbf{D}(2
\mathbf{B}(t) \langle \mathbf{r} \rangle (t)+ \boldsymbol{\Delta}(t)) \, , \qquad
\frac{d}{dt} \boldsymbol{\sigma}(t)=2 i (\boldsymbol{\sigma}(t)
\mathbf{B}(t) \mathbf{D}- \mathbf{D} \mathbf{B}(t) \boldsymbol{\sigma}(t)) \, .
\label{dsigma}
\end{equation}
These differential equations, being first obtained in
\cite{vdod1,vdod2} are the generalization of the one-dimensional case discussed
in the previous section. In our case, the nonlinear differential equations for
the density matrix parameters can be obtained by explicit calculation of the
covariances at time $t$. The resulting equations can then be solved by the
substitution of the solution of equation~(\ref{dsigma}) or by direct
integration.

To make the relation easier to see, we point out that the $2 N
\times 2 N$ symplectic matrix $\mathbf{D}$ contains in its
diagonal, blocks made of
the $2 \times 2$ symplectic matrix $\boldsymbol{\Sigma}$, that is,
\[
\mathbf{D}=-i \left(
\begin{array}{ccccc}
 \boldsymbol{\Sigma} & 0 & 0 & \cdots & 0 \\
0 &  \boldsymbol{\Sigma} & 0 & \cdots & 0 \\
\vdots & \ddots & \ddots &  \ddots & \vdots \\
0 & \cdots & 0 &  \boldsymbol{\Sigma} & 0 \\
0 & 0& \cdots & 0 &  \boldsymbol{\Sigma}
\end{array}
\right) \, .
\]
One can also notice that the differential equations for the entries of the
covariance matrix are linear and can be expressed in the following matrix
form:
\begin{equation}
\frac{d}{dt} \mathbf{v}= \mathbf{M} \mathbf{v} \, ,
\label{vector}
\end{equation}
where $\mathbf{v}$ is a $N(2N+1)$-dimensional vector, which contains all the
independent covariances in the $N$-partite system, and the matrix $\mathbf{M}$
is a square matrix of the same dimension that contains the Hamiltonian
coefficients.

\section{Nonunitary Evolution for Gaussian Subsystems}
Assume that the operators in the Hamiltonian of
equation~(\ref{ham-mult}) correspond to the ones of a multipartite
system, where the position and momentum for the $j$th part are given
by $\hat{p}_j$ and $\hat{q}_j$, respectively. Given this, one can
see that the evolution of the complete system is unitary, but each
one of its parts evolves in a nonunitary way due to the correlations
between these parts. When the complete system is Gaussian, each one
of its parts is also Gaussian. To show this property, lets assume
that the $N$-partite system can be determined by the following
density matrix at time $t=0$:
\begin{equation}
\langle \mathbf{x} \vert \hat{\rho}(0) \vert \mathbf{x}' \rangle=N
\exp \left\{ -\frac{1}{2} \tilde{\mathbf{y}}\, \mathbf{A} \,
\mathbf{y}+\tilde{\mathbf{b}}\, \mathbf{y} \right\},
\label{rho0N}
\end{equation}
where $\tilde{\mathbf{x}}=(x_1,x_2,\ldots,x_N)$,
$\tilde{\mathbf{x}}'=(x'_1,x'_2,\ldots,x'_N)$, and
$\tilde{\mathbf{y}}=(x_1,x_2,\ldots,x_N,x'_1,x'_2,\ldots,x'_N)$ are
real vectors. Also, we define the vector
$\tilde{\mathbf{b}}=(b_1,b_2,\ldots,b_N,b_1^*, b_2^*,\ldots,b_N^*)$,
and the matrix
\[
\mathbf{A}=\left(
\begin{array}{cc}
\mathbf{u} & -\mathbf{v} \\
-\tilde{\mathbf{v}} & \mathbf{u}^*
\end{array}
\right)\, ,
\]
where the matrices $\mathbf{u}$ and $\mathbf{v}$ can be written as
\[
\mathbf{u}= \left(
\begin{array}{cccc}
2 a_{1,1} & -a_{1,2} & \cdots & -a_{1,N}\\
-a_{1,2} & 2 a_{2,2} & \cdots & -a_{2,N} \\
\vdots & \vdots & \ddots & \vdots \\
-a_{1,N} & -a_{2,N} &  \cdots & 2 a_{N,N}
\end{array}
\right) \, , \quad
\mathbf{v}=\left(
\begin{array}{ccccc}
 a_{1,N+1} & a_{1,N+2} & \cdots &a_{1,2N-1} & a_{1,2N}\\
a_{1,N+2}^* & a_{2,N+2} & \cdots & a_{2,2N-1} & a_{2,2N} \\
\vdots & \vdots &  \ddots & \vdots & \vdots \\
a_{1,2N-1}^* & a_{2,2N-1}^* &  \cdots &  a_{N-1,2N-1} & a_{N-1,2N}\\
a_{1,2N}^* & a_{2,2N}^* & \cdots & a_{N,2N-1}^* & a_{N,2N}
\end{array}
\right)\, .
\]

It is common knowledge that the dynamics of the composite system is determined
by the evolution of its covariance matrix and the mean values of the position
and momentum operators, i.e., by the solution of equation~(\ref{dsigma}) with
the initial state~(\ref{rho0N}). The resulting state has the same purity as
the initial state, since the evolution is unitary. However, there exists a
nonunitary evolution of the parts, which compose the $N$-partite system.

To obtain the dynamic evolution of one of the parts, we can use the
partial trace method. In other words, one should take the partial
trace of all the subsystems in $\langle \mathbf{x}
\vert\hat{\rho}(t) \vert
\mathbf{x}' \rangle$, except the one we want to study. Nevertheless,
as the system is Gaussian, the partial traces should give us also a
Gaussian state for the density matrix under study.

As the most general one-dimensional Gaussian state can be obtained
by the $2\times 2$ covariance matrix and the mean values ($\langle
\hat{p}
\rangle (t)$ and $\langle \hat{q} \rangle (t)$), we can obtain the
result from the solutions to equation~(\ref{dsigma}) without the
necessity of the partial trace operation.

On the other hand, once the time derivatives of these properties are
established, one can derive the differential equation that the
density matrix for the subsystem must satisfy. To show this
procedure, we can take the bipartite system as an example.

\subsection{Nonunitary Evolution on a Bipartite System}
To exemplify the nonunitary evolution of a subsystem within a
system, one can take a bipartite Gaussian state which evolves on the
Hamiltonian
\begin{equation}
\hat{H}(t)=\tilde{\mathbf{r}}\mathbf{B}(t)\mathbf{r}+\tilde{\boldsymbol{\Gamma}}(t)
\mathbf{r}=(\hat{p}_1, \hat{q}_1,\hat{p}_2,\hat{q}_2)
\left( \begin{array}{cccc}
\omega_{1,1} & \omega_{1,2} & \omega_{1,3} & \omega_{1,4}\\
\omega_{1,2} & \omega_{2,2} & \omega_{2,3} & \omega_{2,4}\\
\omega_{1,3} & \omega_{2,3} & \omega_{3,3} & \omega_{3,4}\\
\omega_{1,4} & \omega_{2,4} & \omega_{3,4} & \omega_{4,4}
\end{array}\right)
\left(\begin{array}{cccc}
\hat{p}_1 \\
\hat{q}_1 \\
\hat{p}_2 \\
\hat{q}_2
\end{array}\right)+ (\gamma_1, \gamma_2,\gamma_3,\gamma_4)
\left(\begin{array}{cccc}
\hat{p}_1 \\
\hat{q}_1 \\
\hat{p}_2 \\
\hat{q}_2
\end{array}\right)\\,
\label{ham_bip}
\end{equation}
where $\omega_{j,k}$ and $\gamma_j$ may be functions of time. In order to
determine the time evolution of the system, one can solve the differential
equations defined for the covariance matrix and the mean values of the
position and momentum operators or, similarly to the one-dimensional case, one
can solve the equations for the density matrix parameters given in the
appendix~(\ref{parameters}). The differential equations for the covariance
matrix and mean values of the position and momentum operators for the
subsystem can be obtained using equation~(\ref{dsigma}). To solve the time
derivative equations, we express the matrices $\mathbf{B}(t)$, $\mathbf{D}$,
and $\boldsymbol{\sigma}(t)$ in the $2\times 2$ block representation; in such
a case, we have
\[
\mathbf{B}(t)=\left(\begin{array}{cc}
\mathbf{B}_1 (t) & \mathbf{B}_{1,2} (t) \\
\tilde{\mathbf{B}}_{1,2} (t) & \mathbf{B}_2 (t)
\end{array}\right) \, , \quad
\mathbf{D}=\left(\begin{array}{cc}
-i \boldsymbol{\Sigma} & 0 \\
0 & -i \boldsymbol{\Sigma}
\end{array}\right) \, , \quad
\boldsymbol{\sigma}(t)=\left(\begin{array}{cc}
\boldsymbol{\sigma}_1(t) & \boldsymbol{\sigma}_{1,2}(t)\\
\tilde{\boldsymbol{\sigma}}_{1,2}(t) & \boldsymbol{\sigma}_2(t)
\end{array}\right) \, ,
\]
where $\boldsymbol{\sigma}_1(t)$ and $\boldsymbol{\sigma}_2(t)$ are the
covariance matrices for the subsystems $1$ and $2$, respectively, and
$\boldsymbol{\sigma}_{1,2}(t)$ is a matrix containing the covariances
associated to the correlations between the two subsystems. The same can be
said for the matrix linked to the Hamiltonian~(\ref{ham_bip}), i.e.,
$\mathbf{B}(t)$ where the block matrices $\mathbf{B}_1(t)$ and
$\mathbf{B}_2(t)$ are associated to the subsystems 1 and 2, respectively,
while $\mathbf{B}_{1,2}$ is associated to the interactions between these two
subsystems.

After this identification, the expression for the covariance
matrices of the subsystems and the correlations can be given as
follows:
\begin{eqnarray}
\dot{\boldsymbol{\sigma}}_1(t)=2\left((\boldsymbol{\sigma}_1(t)
\mathbf{B}_1 (t)+\boldsymbol{\sigma}_{1,2}(t)\tilde{\mathbf{B}}_{1,2})
\boldsymbol{\Sigma}-\boldsymbol{\Sigma}(\mathbf{B}_1 (t) \boldsymbol{\sigma}_1(t)
+\mathbf{B}_{1,2} \tilde{\boldsymbol{\sigma}}_{1,2} (t))\right) \, , \nonumber \\
\dot{\boldsymbol{\sigma}}_2(t)=2\left((\boldsymbol{\sigma}_2(t) \mathbf{B}_2 (t)
+\tilde{\boldsymbol{\sigma}}_{1,2}(t)\mathbf{B}_{1,2})\boldsymbol{\Sigma}
-\boldsymbol{\Sigma}(\mathbf{B}_2 (t) \boldsymbol{\sigma}_2(t)+\tilde{\mathbf{B}}_{1,2}
\boldsymbol{\sigma}_{1,2} (t))\right) \, ,  \\
\dot{\boldsymbol{\sigma}}_{1,2}(t)=2\left((\boldsymbol{\sigma}_1(t)
\mathbf{B}_{1,2} (t)+\boldsymbol{\sigma}_{1,2}(t)\mathbf{B}_2)\boldsymbol{\Sigma}
-\boldsymbol{\Sigma}(\mathbf{B}_1 (t)
\boldsymbol{\sigma}_{1,2}(t)+\mathbf{B}_{1,2} \boldsymbol{\sigma}_2
(t))\right) \, .\nonumber
\end{eqnarray}
Then one can recognize the term $2(\boldsymbol{\sigma}_j
\mathbf{B}_j\boldsymbol{\Sigma}-\boldsymbol{\Sigma}\mathbf{B}_j\boldsymbol{\sigma}_j)$
for $j=1,2$, as the term corresponding to a unitary evolution of each
subsystem~(\ref{covg}). The extra term
$2(\boldsymbol{\sigma}_{1,2}\tilde{\mathbf{B}}_{1,2}\boldsymbol{\Sigma}
-\boldsymbol{\Sigma}\mathbf{B}_{1,2}\tilde{\boldsymbol{\sigma}}_{1,2})$ is
associated to the nonunitary evolution of the subsystems.

It is worth noting that these results are in accordance with the ones
described by Sandulescu et al.~\cite{sandulescu} and Isar
\cite{isar09,isar18}, where those results are obtained by solving the
Gorini--Kossakowski--Sudarshan--Lindblad master
equation~\cite{kossa,lind,gori} for two coupled oscillators. The main
difference here is that our results are obtained exactly from the von Neumann
equation without introducing a master equation.

\subsection{Invariant and Quasi-Invariant States}
The expression for the derivatives of the covariance matrix can lead to the
definition of different Gaussian states, which do not evolve in the
Hamiltonian dynamics. These type of states can be found as solutions to the
equation $\dot{\boldsymbol{\sigma}}=0$, which can be expressed in terms of the
following equation regarding the covariance and the Hamiltonian matrices
$\boldsymbol{\sigma}\mathbf{B}(t)\mathbf{D}-\mathbf{D}\mathbf{B}(t)\boldsymbol{\sigma}=0$.
As discussed before, this system of differential equations can be replaced by
$\dot{\mathbf{v}}=\mathbf{Mv}$~(\ref{vector}) with the following
correspondences:
\[
\tilde{\mathbf{v}}=(\sigma_{p_1p_1},\sigma_{p_1q_1},\sigma_{p_1p_2},
\sigma_{p_1q_2},\sigma_{q_1q_1},\sigma_{q_1p_2},\sigma_{q_1q_2},
\sigma_{p_2p_2},\sigma_{p_2q_2},\sigma_{q_2q_2})\, , \nonumber \\
\]
and the matrix $\mathbf{M}$ containing the Hamiltonian coefficients
is presented in equation (\ref{eme2}) of Appendix~B. It is possible to see that the matrix $\mathbf{M}$ has a determinant
$\det\, \mathbf{M}=0$ and a rank $\mathcal{R}=8$. From these
properties, one can conclude that the system
$\dot{\boldsymbol{\sigma}}(t)=\dot{\mathbf{v}}=0$ has at most two
different nontrivial solutions which may be physical.

To exemplify the definition of bipartite states, which have
stationary behavior, we consider the frequency converter and the
parametric amplifier. Both of these systems are quadratic and model
the interaction between different electromagnetic fields on a
nonlinear medium.

\subsection{Frequency Converter}
The quantum frequency converter is a device, where two different
unimodal electromagnetic fields, called the input and the output,
interact with a semiclassical pump field on a nonlinear material.
This interaction has the goal to interchange the frequencies of the
input and output beams at specific times. This behavior can be
modeled using the following Hamiltonian:
\begin{equation}
\hat{H}(t)=\hbar \omega_1 \left(\hat{a}_1^\dagger \hat{a}_1+ {1}/{2}
\right)+\hbar \omega_2 \left(\hat{a}_2^\dagger \hat{a}_2+{1}/{2} \right)-\hbar
\kappa \left(\hat{a}_1^\dagger \hat{a}_2 e^{-i \omega t}+\hat{a}_1
\hat{a}_2^\dagger e^{i \omega t}\right) \, ,
\end{equation}
where the frequencies $\omega_{1,2}$ are the input and output frequencies,
respectively, $\omega$ is the pump field frequency, and the bosonic operators
$\hat{a}_{1,2}$ are the annihilation operators of the input and output fields,
respectively. These beams interact with an intensity $\kappa$ in a nonlinear
medium as, e.g., a nonlinear crystal. In this case, the Hamiltonian matrix
$\mathbf{B}(t)$ from~(\ref{ham-mult}) (in a unit system where $\hbar=1$) reads
\[
\mathbf{B}(t)=\frac{1}{2}\left(
\begin{array}{cccc}
1 & 0 & -\dfrac{\kappa}{\sqrt{\omega_1 \omega_2}} \cos (\omega t) & \kappa \sqrt{{\omega_2}/{\omega_1}} \sin (\omega t) \\
0 & \omega_1^2 & -\kappa \sqrt{{\omega_1}/{\omega_2}} \sin (\omega t) & -\kappa \sqrt{\omega_1 \omega_2} \cos (\omega t) \\
-\dfrac{\kappa}{\sqrt{\omega_1 \omega_2}} \cos (\omega t) &  -\kappa \sqrt{{\omega_1}/{\omega_2}} \sin (\omega t) & 1 & 0 \\
 \kappa \sqrt{{\omega_2}/{\omega_1}} \sin (\omega t) & - \kappa \sqrt{\omega_1 \omega_2} \cos (\omega t) & 0 & \omega_2^2
\end{array}
\right) \, .
\]
For this Hamiltonian, one can obtain different states that have
dynamical equilibrium properties, i.e., states with a time
derivative for the covariance matrix equal to zero. To characterize
these type of states, one should solve the equation~(\ref{dsigma}) with
$\dot{\boldsymbol{\sigma}} (t)=0$. As previously discussed,
$\dot{\boldsymbol{\sigma}}(t)=0$ can be expressed in vector form as
\begin{equation}
\mathbf{Mv}=0 \, ,
\label{emeve}
\end{equation}
where $\mathbf{v}$ is defined as
\[
\mathbf{v}=\left(\sigma_{p_1 p_1}, \sigma_{p_1 q_1},\sigma_{p_1 p_2},
\sigma_{p_1 q_2}, \sigma_{q_1 q_1}, \sigma_{q_1 p_2}, \sigma_{q_1 q_2},
\sigma_{p_2 p_2}, \sigma_{p_2 q_2}, \sigma_{q_2 q_2} \right) \, ,
\]
and the matrix $\mathbf{M}$ is a $10 \times 10$ matrix of rank
$\mathcal{R}=8$, which contains the Hamiltonian parameters of $\mathbf{B}(t)$.
To solve equation~(\ref{emeve}), one can explore the null space of matrix
$\mathbf{M}$. The resulting null space contains two different vectors, one
contains a not physical solution. In this solution,
\begin{eqnarray}
\sigma_{p_1p_1}=(\omega_1^{3}\omega_2)^{1/2}(\omega_2-\omega_1)\sec (\omega t)/\kappa, &\quad&  \sigma_{p_1p_2}=\omega_1 \omega_2, \nonumber \\
 \sigma_{p_1 q_2}=-\omega_1 \tan (\omega t),&\quad& \sigma_{q_1 q_1}=\omega_2^{1/2}(\omega_2-\omega_1)\sec (\omega t)/(\kappa \omega_1^{1/2}),\nonumber \\
  \sigma_{q_1 p_2}=\omega_2 \tan (\omega t), &\quad& \sigma_{q_1 q_2}=1, \nonumber
\end{eqnarray}
while all the other covariances are equal to zero. In particular, it contains
the nonphysical terms $\sigma_{p_2 p_2}=\sigma_{q_2 q_2}=0$ that resembles the
case where one of the subsystem is classical, as in a classical system the
values of the covariances can be equal to zero. The null space also contains
another vector which has the following physical values:
\begin{equation}
\sigma_{p_1 p_1}= \omega_1 \omega_2,\quad \sigma_{q_1 q_1}=
{\omega_2}/{\omega_1},\quad \sigma_{p_2 p_2}=\omega_2^2,\quad
\sigma_{q_2 q_2}=1 \, ,
\label{signull}
\end{equation}
with the remaining covariances equal to zero.

These results led us to the conclusion that a two-mode Gaussian state with
initial covariances proportional to the ones established in~(\ref{signull})
have the same covariances for any time $t>0$ (for time-independent
parameters $\omega$, $\omega_{1,2}$, and $\kappa$). This property has several
physical implications such as, for example, that the purity of the subsystems will
be always the same regardless of the interaction between them and despite the
interchange of their frequencies. The resulting states will only have
different mean values of the quadrature components ($p_1$, $q_1$, $p_2$, and
$q_2$), which evolve according to the classical Hamilton equations.

In the case where the mean values of the quadrature components
$b_j$; $j=1,2$ are equal to zero  in equation~(\ref{rho0N}), one can
obtain different states, which do not change their properties over time.
In this case, the entanglement of the system (which can be obtained
by the logarithmic negativity of the covariance matrix) will also be
an invariant of the system. These properties make the evolution of
this type of states relevant to quantum computing and quantum
information.

By the use of the inverse relations of equations~(\ref{ap1}) and (\ref{ap2}),
one can then write a general state with invariant covariance matrix, which
only changes its mean values according to the classical motion equations. Such
state can be expressed as the one in~(\ref{rho0N}), after making the
identification
\begin{eqnarray}
\mathbf{A}=\left(\begin{array}{cccc}
\dfrac{\omega_1}{4 C \omega_2}+C \omega_1 \omega_2 & 0 & \dfrac{\omega_1}{4 C \omega_2}- C \omega_1 \omega_2 & 0 \\
0 & \dfrac{1}{4 C}+C \omega_2^2 & 0 & \dfrac{1}{4 C}- C \omega_2^2 \\
\dfrac{\omega_1}{4 C \omega_2}- C \omega_1 \omega_2 & 0 & \dfrac{\omega_1}{4 C \omega_2}+C \omega_1 \omega_2 \\
0 & \dfrac{1}{4 C}- C \omega_2^2 & 0 & \dfrac{1}{4 C}+C \omega_2^2
\end{array}
\right) \, ,
\end{eqnarray}
with $C$ being a constant, which needs to be chosen in order for
the covariance matrix to be positive. In particular, to fulfill the Schr\"odinger--Robertson inequalities
$\sigma_{p_i p_i} \sigma_{q_i q_i}-\sigma_{p_i q_i}^2\geq 1/4$ ($i=1,2$) and
$\det \boldsymbol{\sigma}\geq1/16$.

\subsection{Parametric Amplifier}

The other Hamiltonian, which can be taken as an example, is the
nondegenerated parametric amplifier. This system also describes the
interaction of an input and output beams with the pump field in a
nonlinear medium. As a result of this interaction, one can obtain
the amplification of the input beam. The Hamiltonian associated to
the parametric amplifier reads
\begin{equation}
\hat{H}=\hbar \omega_1 \left( \hat{a}_1^\dagger \hat{a}_1+{1}/{2} \right)
+\hbar \omega_2 \left( \hat{a}_2^\dagger \hat{a}_2+{1}/{2} \right) -\hbar
\kappa \left(\hat{a}_1^\dagger \hat{a}_2^\dagger e^{-i \omega t}+\hat{a}_1
\hat{a}_2 e^{i \omega t}\right) \, ,
\end{equation}
where the frequencies $\omega_{1,2}$ are the frequencies of the
input and output beam channels, and $\omega$ is the frequency of a
pump field which allows the amplification of the input channel. Then
the Hamiltonian matrix $\mathbf{B}(t)$ is
\begin{equation}
\mathbf{B}(t)=
\frac{1}{2}\left( \begin{array}{cccc}
1 & 0 & \kappa \dfrac{\cos(\omega t)}{\sqrt{\omega_1 \omega_2}} & \kappa \sqrt{{\omega_2}/{\omega_1}} \sin(\omega t) \\
0 & \omega_1^2 & \kappa \sqrt{{\omega_1}/{\omega_2}} \sin(\omega t) & -\kappa\sqrt{\omega_1 \omega_2} \cos(\omega t) \\
\kappa\dfrac{\cos(\omega t)}{\sqrt{\omega_1 \omega_2}} & \kappa \sqrt{{\omega_1}/{\omega_2}} \sin(\omega t) & 1 & 0 \\
\kappa \sqrt{{\omega_2}/{\omega_1}} \sin(\omega t) & -\kappa\sqrt{\omega_1 \omega_2} \cos(\omega t) & 0 & \omega_2^2
\end{array}
\right) \, .
\end{equation}

\begin{figure}
\begin{subfigure}[]
{\includegraphics[scale=0.28]{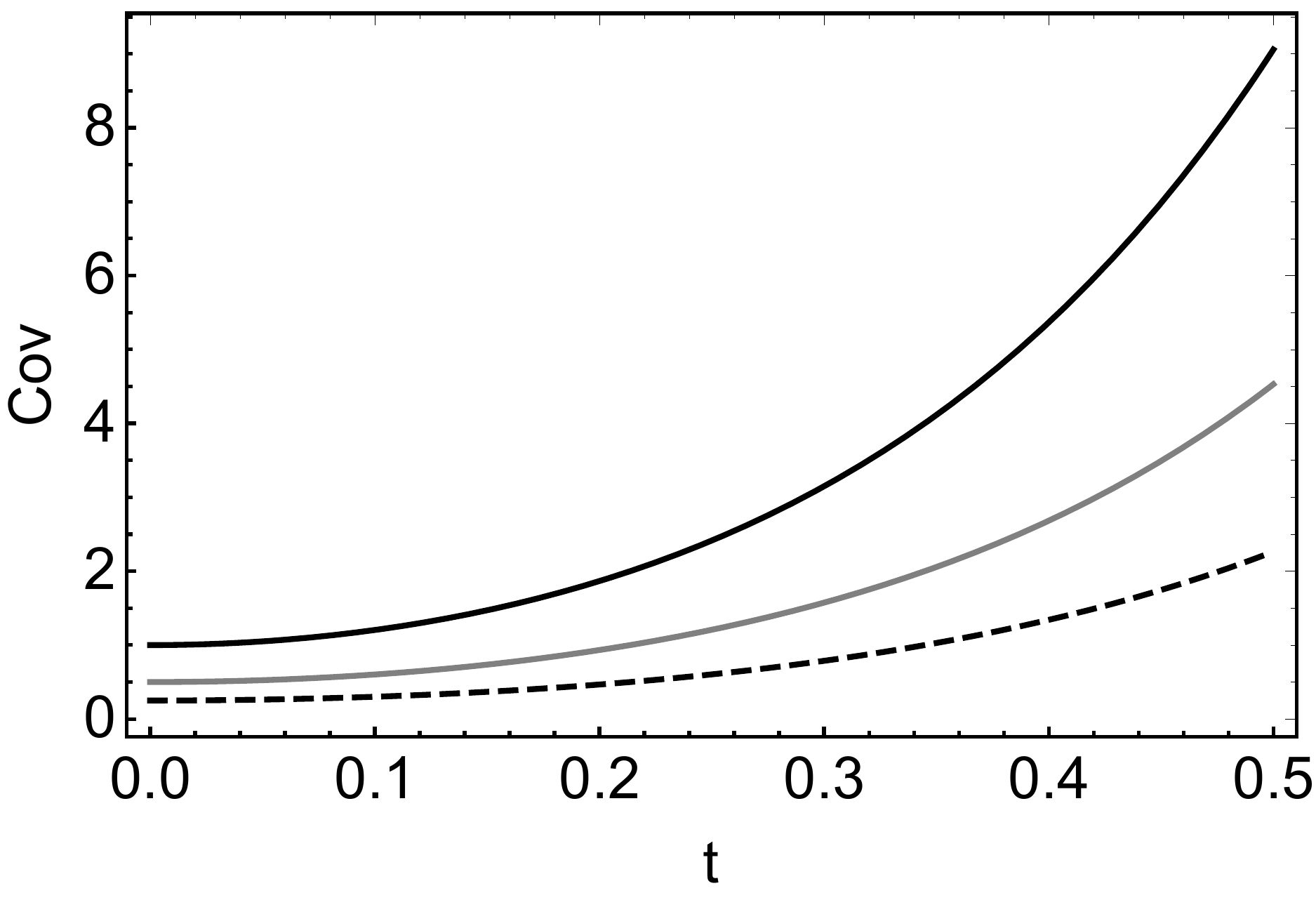} \label{fig:1a}}
\end{subfigure}
\begin{subfigure}[]
{\includegraphics[scale=0.28]{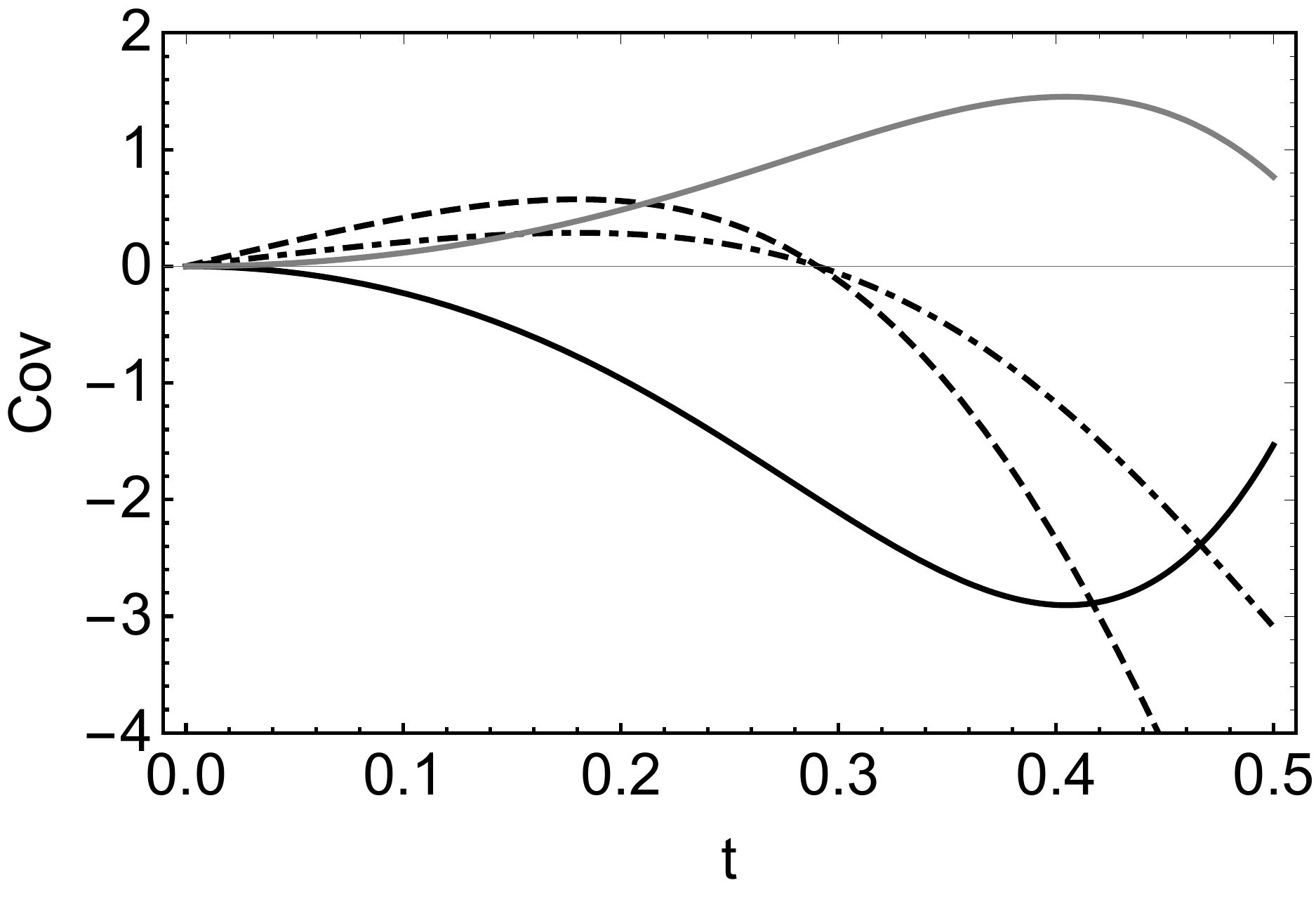} \label{fig:1b}}
\end{subfigure}
\begin{subfigure}[]
{\includegraphics[scale=0.28]{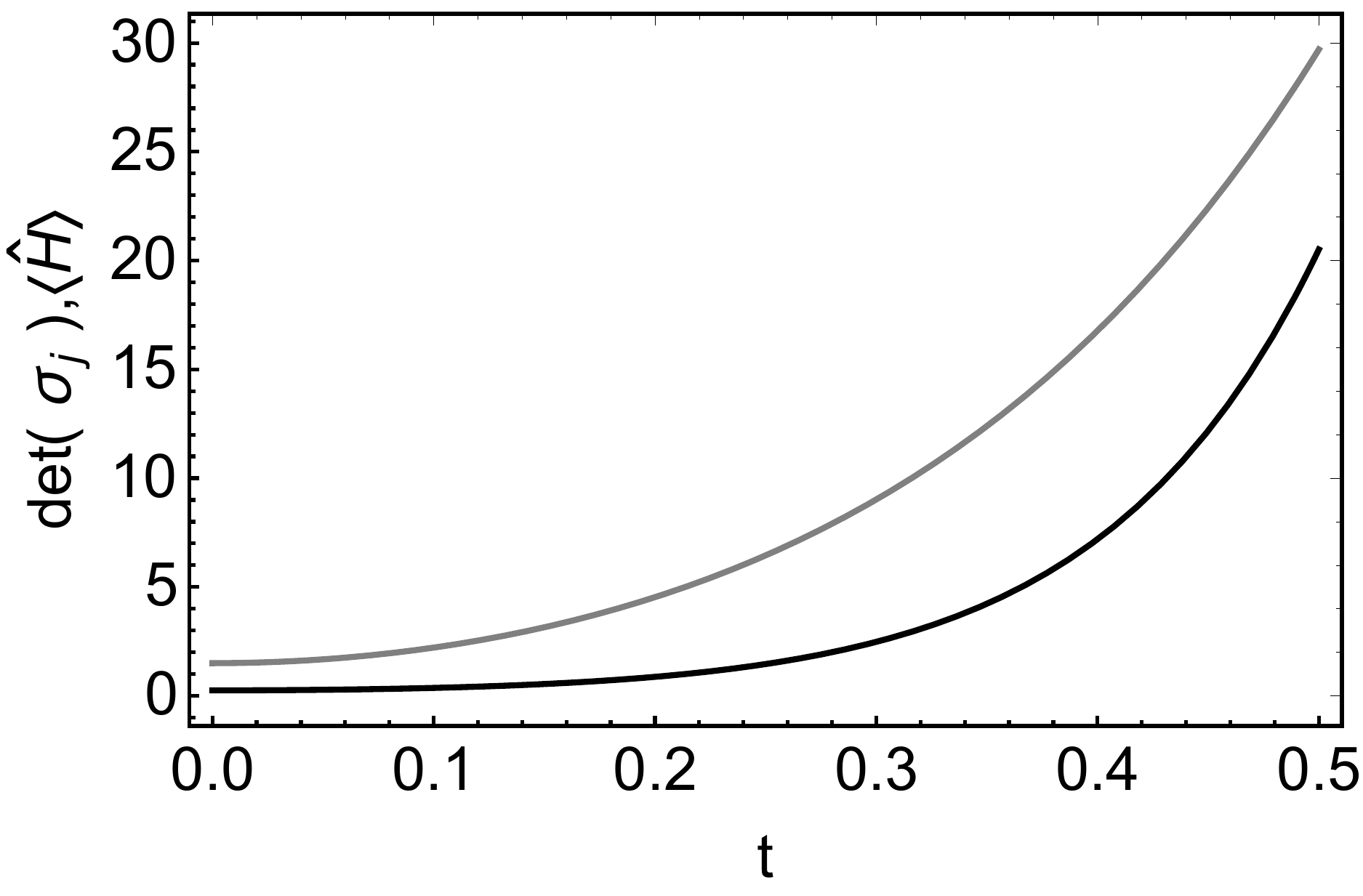} \label{fig:1c}}
\end{subfigure}
\caption{Time evolution for the covariances (a) $\sigma_{p_1 p_1}$
(black), $\sigma_{q_1 q_1}$ (dashed), and $\sigma_{p_2 p_2}=\sigma_{q_2 q_2}$
(gray), (b) the covariances $\sigma_{p_1 p_2}$ (black), $\sigma_{p_1 q_2}$
(black dashed), $\sigma_{p_2 q_1}$ (black dot-dashed), and $\sigma_{q_1 q_2}$
(gray), (c) $\det \boldsymbol{\sigma}_1=\det \boldsymbol{\sigma}_2$ for
the subsystems (black) and the time dependence of the mean value $\langle
\hat{H} \rangle(t)$ (gray). For all the plots, the initial values are
$\sigma_{p_1p_1}(0)=1$, $\sigma_{q_1 q_1}(0)=1/4$, and $\sigma_{p_2
p_2}(0)=\sigma_{q_2 q_2}(0)=1/2$. All the other initial covariances are equal
to zero. The frequencies used are $\omega_1=2$, $\omega_2=1$, $\omega=7$, and
$\kappa=\sqrt{10}$ \label{ampl}}
\end{figure}

Following an analogous procedure to obtain the solutions of the equation
$\dot{\boldsymbol{\sigma}}=0$, one can show that the null space of the
corresponding matrix $\mathbf{M}$ for this problem can lead us to nonphysical
values for different covariances on the system. One of the vectors
of the null space for the case $\omega_{1,2}>0$ can be written as
\begin{eqnarray*}
\sigma_{p_1 p_1}&=&\frac{\omega_1 \sqrt{\omega_1 \omega_2} (\omega_1+\omega_2)
\sec (\omega t)}{\kappa}, \quad \sigma_{p_1 p_2}=-\omega_1 \omega_2, \quad
\sigma_{p_1 q_2}= -\omega_1 \tan (\omega t), \nonumber \\
\sigma_{q_1 q_1}&=& \frac{\omega_2 (\omega_1+\omega_2)
\sec (\omega t)}{\kappa \sqrt{\omega_1 \omega_2}}, \quad
\sigma_{p_2 q_1}=-\omega_2 \tan (\omega t), \quad \sigma_{q_1 q_2}=1 ,
\end{eqnarray*}
while all the other covariances are equal to zero. The other vector on the
null space is
\[
\sigma_{p_1 p_1}=-\omega_1 \omega_2, \quad \sigma_{q_1 q_1}=- \frac{\omega_2}{\omega_1} ,
\quad \sigma_{p_2 p_2}=\omega_2^2, \quad \sigma_{q_2 q_2}=1.
\]
As the condition $\omega_{1,2}>0$ was used to obtain these results then both vectors lead to nonphysical covariances. Nevertheless, one can obtain states with slow change
ratio of the covariances compared with the change of the mean value of the
system Hamiltonian $\langle \hat{H} \rangle (t)$. These type of states can be
defined by considering the initial covariances equal to $C=1/(\omega_1
\omega_2)$ times the ones presented in equation (\ref{signull}); in other
words,
\[
\sigma_{p_1,p_1}=1, \quad \sigma_{q_1,q_1}=1/\omega_1^2, \quad \sigma_{p_2,p_2}=\omega_2/\omega_1, \quad \sigma_{q_2,q_2}=1/(\omega_1 \omega_2) \, .
\]
The slow time-dependence behavior of these covariances can be seen in Fig.~(\ref{ampl}), where the
time dependence of the covariances and the purity of the subsystems are
illustrated. The evolution of the subsystems in the parametric amplifier
normally varies very fast, as the photons from the pump field are transformed
in photons of both subsystems. Nevertheless, it can be seen in
Fig.~(\ref{ampl}) that the variation of the majority of the covariances is not
as fast compared with the change of $\langle \hat{H}\rangle(t)$,
providing the strong coupling between the subsystems. In this particular
example, one can see that $\sigma_{p_2 p_2}(t)=\sigma_{q_2 q_2}(t)$ and
$\sigma_{p_1 q_1}(t)=\sigma_{p_2 q_2}(t)=0$.

The detection of these type of states can be done by the use of quantum
tomography, as it is discussed in the next section.

\section{Gaussian States and Their Evolution in the Tomographic-Probability
Representation}

There exist different representations of quantum
states~\cite{Ramon-JRLR,Entropy-2-692-2018,Milestones,Roberto,PLA2020,Julio},
between them the probability tomographic representation is of particular
interest. In this representation, e.g., one-mode photon states are identified
with symplectic tomograms~\cite{TombesiPLA}, which correspond to the
conditional probability distribution $w(X\mid\mu,\nu)$ of the photon
quadrature $-\infty<X<\infty$, to be measured in a reference frame with
parameters $\mu=s\cos\theta$ and $\nu=s^{-1}\sin\theta$.
Here,$-\infty<\mu,\nu<\infty$, $s$ is a time scaling parameter, and $\theta$
is the local oscillator phase, which is used in experiments~\cite{Raymer1993}
to obtain the Wigner function of the photon state.

The symplectic tomogram $w_\rho(X\mid\mu,\nu)$ is determined by the photon
density operator $\hat\rho$~\cite{TombesiPLA} as
\begin{equation}\label{TPR1}
w_\rho(X\mid\mu,\nu)=\mbox{Tr} \left[\hat\rho\,\delta(X\hat 1-\mu\hat
q-\nu\hat p)\right],
\end{equation}
where $\hat q$ and $\hat p$ are quadrature components -- the position and
momentum operators within the framework of the oscillator model of the
one-mode electromagnetic-field photons. The symplectic tomogram satisfies the
normalization condition
\begin{equation}\label{TPR2}
\int w_\rho(X\mid\mu,\nu)\,dX=1,
\end{equation}
and inversely, it determines the density operator $\hat\rho$ of the photon state
\begin{equation}\label{TPR3}
\hat\rho=\frac{1}{2\pi}\int w(X\mid\mu,\nu)\,\exp\left[i\,\delta(X\hat
1-\mu\hat q-\nu\hat p)\right]dX\,d\mu\,d\nu.
\end{equation}

The optical tomogram of the photon state $w_{\rm opt}(X\mid\theta)\equiv
w(X\mid\mu=\cos\theta,\nu=\sin\theta)$ is measured in experiments and, in view
of the homogeneity property of the Dirac delta-function, the measured optical
tomogram of the photon state determines the symplectic tomogram
\begin{equation}\label{TPR4}
w(X\mid\mu,\nu)=\frac{1}{\sqrt{\mu^2+\nu^2}}\,w_{\rm opt}\left[
\frac{X}{\sqrt{\mu^2+\nu^2}}\Bigg\vert\arctan\frac{\nu}{\mu}\right].
\end{equation}
For Gaussian states~(\ref{gaus0}), the tomographic-probability distribution of
random photon quadrature $X$ has the conventional form of normal distribution
\begin{equation}\label{TPR5}
w(X\mid\mu,\nu)=\frac{1}{\sqrt{2\pi\sigma(\mu,\nu)}}\,\exp\left[
-\frac{\left(X-\bar X(\mu,\nu)\right)^2}{2\sigma(\mu,\nu)}\right].
\end{equation}
In view of (\ref{TPR1}), one has the mean value of the quadrature
\begin{equation}\label{TPR6}
\bar X(\mu,\nu)=\mu\langle\hat q\rangle + \nu\langle\hat p\rangle
\end{equation}
and the covariance of the quadrature $\sigma(\mu,\nu)$ reads
\begin{equation}\label{TPR7}
\sigma(\mu,\nu)=\mu^2\sigma_{qq} + \nu^2\sigma_{pp} + 2\mu\nu\sigma_{pq}.
\end{equation}
For measured optical tomogram, the dispersion $\sigma(\theta)$ is
\begin{equation}\label{TPR8}
\sigma(\theta)=(\cos^2\theta)\sigma_{qq} + (\sin^2\theta)\sigma_{pp} + (\sin
2\theta)\sigma_{pq}.
\end{equation}
In the quantum system with Hamiltonian~(\ref{ham_ex1d}), the tomographic
quadrature dispersion~(\ref{TPR7}) evolves according to the evolution
\begin{equation}\label{TPR9}
\sigma(\mu,\nu,t)=\mu^2\sigma_{qq}(t) + \nu^2\sigma_{pp}(t)  +
2\mu\nu\sigma_{pq}(t),
\end{equation}
where $\sigma_{qq}(t)$, $\sigma_{pp}(t)$, and $\sigma_{pq}(t)$ are provided by
explicit expressions~(\ref{sol_ex1d}) and parameters $\langle\hat p(t)\rangle$
and $\langle\hat q(t)\rangle$ are given by~(\ref{mean_ex1d}). Thus, the properties of Gaussian states of oscillator with time-dependent
parameters described by the covariances of the position and
momentum and their mean values can be checked by considering the covariance of the homodyne quadrature $X$, as well as the mean value
evolution.

The properties of the invariant Gaussian states can be visualized by the
properties of either the Wigner function or the tomographic-probability
distributions. There are examples of the time-dependent Gaussian-packet solutions of
the kinetic equation for the sympectic tomogram~\cite{TombesiFP,KorennoyJRLR} in
the case of harmonic oscillator Hamiltonian~(\ref{ham_ex1d}) with $\nu=0$. Since the
dispersion matrix for the quadrature $X$ is the linear combination of
covariances $\sigma_{qq}(t)$, $\sigma_{pp}(t)$, and
$\sigma_{pq}(t)$ which, in the case of invariant Gaussian states, do not
depend on time, the state tomogram also does not depend on time. The invariant states with density operators $\mid E_n\rangle\langle E_n\mid$
have the oscillator tomograms obtained from energy states $\mid E_n\rangle$,
where $\hat H\mid E_n\rangle=E_n\mid E_n\rangle$. Tomograms of invariant
Gaussian states satisfy the equality
\begin{equation}\label{TPR10}
{\cal P}_G^n=\frac{1}{2\pi}\int w_G(X\mid\mu,\nu)\,w_{E_n}(Y\mid\mu,\nu)
\,e^{i(X+Y)}\,dX\,dY\,d\mu\,d\nu,
\end{equation}
where the parameter ${\cal P}_G^n$ is the probability to detect the properties of
the stationary state $\mid E_n\rangle$ with energy value $E_n$ in the Gaussian
state with the time-dependent tomogram $w_G(X\mid\mu,\nu)$. This state also
does not depend on time.

Any convex sum of states $\mid E_n\rangle\langle E_n\mid$ is a density
operator. One can conjecture that there is the decomposition of normal
distribution $w_G(X\mid\mu,\nu)$ corresponding, e.g., to a thermal state with
$\hat{\rho}=\exp(- \hat{H}/(kT))/{\rm Tr} (\exp(- \hat{H}/(kT)))$ ($T$ being
the temperature and $k$ is the Boltzmann constant), which can be presented as
\begin{equation}\label{TPR11}
w_G(X\mid\mu,\nu)=\sum_n{\cal P}_G^n\,w_{E_n}(X\mid\mu,\nu),\qquad\sum_n{\cal
P}_G^n=1.
\end{equation}
An analogous relation can be then written also for the Wigner function of the
invariant Gaussian state of the oscillator, as well as for the density matrix in
the position representation.

Now we consider a harmonic oscillator with the Hamiltonian $\hat H=\dfrac{\hat
p^2}{2}+\dfrac{\hat q^2}{2}$. The density matrix of thermal equilibrium state
with temperature $T=\beta^{-1}$ in the position representation reads (here, we
assume $\hbar=\omega=m=k=1$)
\begin{equation}\label{TPR12}
\rho(x,x',\beta)=\left[\pi^{-1}\tan^2(\beta/2)\right]^{1/2}\exp\left[\frac{xx'}{\sinh\beta}
-\frac{x^2+x'^2}{2}\coth\beta\right].
\end{equation}
Green function of the oscillator has the Gaussian form
\begin{equation}\label{TPR13}
G(x,x',t)=\langle x\mid e^{-it\hat H}\mid x'\rangle =\frac{1}{\sqrt{2\pi i\sin
t}}\exp\left[\frac{i(x^2+x'^2)}{2}\cot t-\frac{ixx'}{\sin t} \right].
\end{equation}
Since the density matrix~(\ref{TPR12}) is determined by the Green
function~(\ref{TPR13}), i.e.,
\begin{equation}\label{TPR14}
\rho(x,x',\beta)=\frac{G(x,x',-i\beta)}{Z(\beta)},
\end{equation}
with the partition function $Z(\beta)$ given by the formula
\begin{equation}\label{TPR15}
Z(\beta)=\sum_{n=0}^\infty\mbox{Tr}\left[\exp\left(-\beta\hat H\right)\mid
n\rangle \langle n\mid\right]=\frac{1}{2\sinh(\beta/2)}\,;
\end{equation}
here, we use the property $\hat H\mid n\rangle=\left(n+1/2\right)\mid
n\rangle$. The density matrix~(\ref{TPR12}) does
not depend on time; this means that in all other representations, as the Wigner
function or tomographic-probability representation, it is time-invariant. The
density matrices of Fock states $\mid n\rangle \langle n\mid$ do not depend on
time.

In the position representation, the density matrix of Fock state $\mid
n\rangle\langle n\mid$ reads
\begin{equation}\label{TPR16}
\langle x\mid n\rangle \langle n\mid x'\rangle
=\frac{H_n(x)H_n(x')}{2^nn!\sqrt\pi}\,\exp
\left(-\frac{x^2}{2}-\frac{x'^2}{2}\right),
\end{equation}
and it does not depend on time.

The density matrix~(\ref{TPR12}), being described by invariant Gaussian
function, is the convex sum of the Fock-state density matrices. One has the
relation
\begin{equation}\label{TPR17}
\rho(x,x',\beta)=\frac{1}{\sqrt\pi}\,\exp
\left(-\frac{x^2}{2}-\frac{x'^2}{2}\right)\sum_{n=0}^\infty
\frac{e^{-(n+1/2)\beta}}{Z(\beta)2^nn!}\,H_n(x)H_n(x'),
\end{equation}
where $Z(\beta)$ is given~(\ref{TPR15}).

In the tomographic-probability representation of the thermal equilibrium
oscillator and Fock states, we have tomograms in explicit form. For Fock states,
\begin{equation}\label{TPR18}
w_n(X\mid\mu,\nu,\beta)=\left[\pi(\mu^2+\nu^2)\right]^{-1/2}\frac{1}{2^nn!}\,
\exp \left(-\frac{X^2}{\mu^2+\nu^2}\right)\,
H_n^2\left(\frac{X}{\sqrt{\mu^2+\nu^2}}\right).
\end{equation}
With all these properties, one can check that the thermal-equilibrium Gaussian
state of equation~(\ref{TPR14}) has a symplectic tomogram in the form of the normal
distribution
\begin{equation}\label{TPR19}
w(X\mid\mu,\nu,\beta)=\frac{1}{\sqrt{2\pi\sigma(\mu,\nu)}} \, \exp
\left(-\frac{X^2}{2\sigma(\mu,\nu)}\right).
\end{equation}

The dispersion of quadrature $\langle X^2\rangle=\sigma(\mu,\nu)$ given by
equation~(\ref{TPR19}) reads
\begin{equation}\label{TPR20}
\sigma(\mu,\nu)=\mu^2\langle\hat q^2\rangle+\nu^2\langle\hat p^2\rangle,
\end{equation}
where the state with density matrix~(\ref{TPR12})
\begin{equation}\label{TPR21}
\langle\hat q^2\rangle=\langle\hat p^2\rangle=\frac{1}{2}\,\coth^2(\beta/2).
\end{equation}
Thus, the symplectic tomogram~(\ref{TPR19}) is given by an invariant normal
probability distribution
\begin{equation}\label{TPR20}
w(X\mid\mu,\nu,\beta)=\frac{\coth(\beta/2)}{\sqrt{\pi(\mu^2+\nu^2)}} \, \exp
\left(-\frac{X^2}{\mu^2+\nu^2}\,\coth^2(\beta/2)\right).
\end{equation}

Since for optical tomogram  $\mu=\cos\theta$, $\nu=\sin\theta$, and
$\mu^2+\nu^2=1$, in the case of thermal single-mode photon state, its optical
tomogram is
\begin{equation}\label{TPR20}
w(X\mid\theta,\beta)=\frac{\coth (\beta/2)}{\sqrt{\pi}} \, \exp
\left(-X^2\,\coth^2(\beta/2)\right);
\end{equation}
it depends neither on the local oscillator phase nor on time. This type of
states are Gaussian and time-independent, so there is a connection between
them and the invariant states discussed above. This connection can be checked
by equaling the covariance matrices in both cases, which can be also checked
experimentally, for example, by preparing an initial Gaussian state according
to the invariance condition $\dot{\boldsymbol{\sigma}}=0$. As seen in previous
examples, this condition implies a value for the initial covariances in terms
of the Hamiltonian parameters. Then using the tomographic representation
discussed here, the relation of these states with thermal states can be
corroborated. As the result of this comparison, one can obtain certain
thermodynamic properties such as the {\it temperature} of the system. This can
also be extended for the bipartite harmonic oscillator, as seen in the next
section.

\section{Two-Mode Gaussian States in the Tomographic-Probability
Representation}

For two-mode harmonic oscillator, the Gaussian-state tomogram is determined
by normal probability distribution of quadratures $X_1$ and $X_2$; it is
expressed in terms of the state density operator $\hat\rho(1,2)$ as follows:
\begin{equation}\label{TMG1}
w(X_1,X_2\mid\mu_1,\nu_1,\mu_2,\nu_2)=\mbox{Tr}\left[\hat\rho(1,2)\,\delta(X_1
-\mu_1\hat q_1-\nu_1\hat p_1)\,\delta(X_2 -\mu_2\hat q_2-\nu_2\hat
p_2)\right];
\end{equation}
in the case of $\langle\hat q_1\rangle=\langle\hat q_2\rangle=\langle\hat
p_1\rangle=\langle\hat p_2\rangle=0$, it reads
\begin{equation}\label{TMG2}
w(X_1,X_2\mid\mu_1,\nu_1,\mu_2,\nu_2)=\frac{1}{2\pi\sqrt{\det\,\boldsymbol{\sigma}(\mu_1,\mu_2,\nu_1,\nu_2)}}
\,\exp\left[-\frac12\left(\tilde{\mathbf{X}} \boldsymbol{\sigma}^{-1}(\mu_1,\nu_1,\mu_2,\nu_2)\mathbf{X}\right)\right].
\end{equation}
Here, $~ \tilde{\mathbf{X}}=(X_1,X_2)~$ and $~\boldsymbol{\sigma}(\mu_1,\nu_1,\mu_2,\nu_2)=
\left(\begin{array}{cc} \langle X_1^2\rangle &  \langle X_1X_2\rangle\\
\langle X_1X_2\rangle & \langle X_2^2\rangle\end{array}\right),$ with
\begin{eqnarray*}
&&\langle X_1^2\rangle=\mu_1^2\langle\hat q_1^2\rangle+\nu_1^2\langle\hat
p_1^2\rangle+\mu_1\nu_1(\langle\hat q_1\hat p_1\rangle+\langle \hat p_1\hat
q_1\rangle),\quad \langle X_2^2\rangle=\mu_2^2\langle\hat
q_2^2\rangle+\nu_2^2\langle\hat p_2^2\rangle+\mu_2\nu_2(\langle\hat q_2\hat
p_2\rangle+ \langle \hat p_2\hat q_2\rangle),\\
&&\qquad\qquad\qquad\qquad\langle X_1X_2\rangle=\mu_1\mu_2\langle\hat q_1\hat
q_2\rangle+\nu_1\nu_2\langle\hat p_1\hat p_2\rangle+\mu_1\nu_2\langle\hat
q_1\hat p_1\rangle+\mu_2\nu_1\langle\hat q_2\hat p_1\rangle.
\end{eqnarray*}

The inverse transform provides the density operator $\hat\rho(1,2)$ expressed
in terms of the tomographic-probability distribution
\begin{eqnarray}\label{TMG3}
\hat\rho(1,2)&=&\frac{1}{4\pi^2}\int
w\left(X_1,X_2\mid\mu_1,\nu_1,\mu_2,\nu_2\right)\nonumber\\
&&\times \exp \left[i\left(X_1+X_2-\mu_1\hat q_1-\nu_1\hat p_1-\mu_2\hat
q_2-\nu_2\hat p_2\right)\right]\,dX_1 \,dX_2\,d\mu_1 \,d\mu_2\,d\nu_1
\,d\nu_2.
\end{eqnarray}
The subsystem tomogram given by the partial trace of the density operator
$\hat\rho(1)=\mbox{Tr}_2\,\hat\rho(1,2)$ reads
\begin{equation}\label{TMG4}
w_1(X_1\mid\mu_1,\nu_1)=\int
w\left(X_1,X_2\mid\mu_1,\nu_1,\mu_2,\nu_2\right)\,dX_2;
\end{equation}
it is also given by the normal distribution discussed in the previous section.

If the tomogram of the two-mode oscillator state corresponds to the solution
of the time evolution equation with a quadratic Hamiltonian, the unitary
evolution of the system can induce nonunitary evolution of
tomogram~(\ref{TMG4}). These evolutions can be used to obtain the shape of the
invariant states, which we have discussed above using the matrix $\mathbf{M}$
shown in Appendix~B.

\section{Summary and Conclusions}
A differential formalism to obtain the time evolution of a multidimensional,
multipartite Gaussian state is defined and studied. This new formalism uses
the time derivative of the parameters of the continuous variable density
matrix of the system. The general procedure to obtain the time evolution can
be summarized as follows: using the derivative of the covariance matrix for
the Gaussian state and the expressions for the covariances in terms of the
parameters of the density matrix, the differential equations for the
parameters of the density function of the system are obtained. The resulting
nonlinear differential equations can be used to obtain new physical
information of the state instead of the use of the Schr\"odinger equation.

This differential formalism can also be used to describe exactly the
nonunitary evolution of the subsystems of a composite Gaussian
state. As an example, we considered a two-mode Gaussian state and
demonstrate that the resulting derivatives of the covariance
matrices for the subsystems contain unitary and nonunitary terms.

This study also allows us to define invariant states, i.e., states which do
not change their properties over time. To show this, we considered systems of
unimodal and bipartite Gaussian with density matrices in the position
representation and the corresponding tomographic-probability representation.
As explicit examples, we presented the invariant states for the
one-dimensional quadratic Hamiltonian and the invariant states for the
two-mode frequency converter and mentioned the applicability of these type of
states in quantum information and computing. Also, quasi-invariant states for
the two-mode parametric amplifier are presented. We point out that discussed
examples of studying parametric systems can be used to apply the results
associated with the behavior of physical systems like photons in cavities with
time-dependent locations of boundaries to dynamical Casimir effect
(see~\cite{dodcas}) and its analog in superconducting
circuits~\cite{olgam,dodosci}. One can discuss nonunitary evolution of
systems, which have no subsystems, using hidden
correlations~\cite{IJQI-Turino}, which are present in noncomposite systems.

\section*{Acknowledgments}

This work was partially supported by DGAPA-UNAM (under Project IN101619). We
would like to express our special thanks to Prof. Dr. Octavio Casta\~nos,
Prof. Dr. Dieter Schuch, and  M. Sc. Hans Cruz for their comments during the
elaboration of the manuscript.

\appendix

\section{A.~Correspondence between the Gaussian Density Matrix Parameters and the Covariance Matrix}
In this appendix, the expressions of the covariance matrix and the
mean values of the Gaussian system given in Sec. 4 are expressed in
terms of the density matrix parameters. For the bipartite system,
one can obtain the following formulas in terms of the density
operator parameters (\ref{rho0N}) and the three covariances $\sigma_{q_1,q_1}$,
$\sigma_{q_1,q_2}$, and $\sigma_{q_2,q_2}$:
\begin{eqnarray}
\sigma_{p_1,p_1}&=&2 a_{11}-(-2a_{11}+a_{13})^2 \sigma_{q_1,q_1}-
(a_{12}+a_{14})^2 \sigma_{q_2,q_2}+2(2a_{11}-a_{13})(a_{12}+a_{14}) \sigma_{q_1,q_2} \, , \nonumber \\
\sigma_{p_1,q_1}&=&i\{(2a_{11}-a_{13})\sigma_{q_1,q_1}-(a_{12}+a_{14})
\sigma_{q_1,q_2}-1/2 \} \, , \nonumber \\
\sigma_{p_1,p_2}&=&-a_{12}+(2a_{22}-a_{24})(a_{12}+a_{14})\sigma_{q_2,q_2}
+(2a_{11}-a_{13})(a_{12}+a_{14}^*)\sigma_{q_1,q_1} \nonumber \\
&
&-\{(-2a_{22}+a_{24})(-2a_{11}+a_{13})+(a_{12}+a_{14}^*)(a_{12}+a_{14})\}\sigma_{q_1,q_2}
\, ,\nonumber\\[-2mm]
&&\label{ap1}\\[-2mm]
\sigma_{p_1,q_2}&=&i \{ (2a_{11}-a_{13}) \sigma_{q_1,q_2}-(a_{12}+a_{14})\sigma_{q_2,q_2}\} \, , \nonumber \\
\sigma_{q_1,p_2}&=& i \{ (2a_{22}-a_{24})\sigma_{q_1,q_2}-(a_{12}+a_{14}^*)\sigma_{q_1,q_1}\} \, , \nonumber \\
\sigma_{p_2,p_2}&=&2a_{22}-(-2a_{22}+a_{24})^2 \sigma_{q_2,q_2}-(a_{12}+a_{14}^*)^2 \sigma_{q_1,q_1}
-2(-2a_{22}+a_{24})(a_{12}+a_{14}^*)\sigma_{q_1,q_2} \, , \nonumber \\
\sigma_{p_2,q_2}&=&i\{ (2a_{22}-a_{24})\sigma_{q_2,q_2}-(a_{12}+a_{14}^*)\sigma_{q_1,q_2}-1/2\} \, ,
 \nonumber
\end{eqnarray}
where the values of the rest of the covariances read
\begin{eqnarray}
\sigma_{q_1,q_1}&=&\frac{2(a_{22}+a_{22}^*-a_{24})}{4(a_{11} + a_{11}^* - a_{13})(a_{22} + a_{22}^* - a_{24})-(a_{12} + a_{12}^* + a_{14} + a_{14}^*)^2} \, , \nonumber \\
\sigma_{q_1,q_2}&=&\frac{a_{12} + a_{12}^* + a_{14} + a_{14}^*}{4(a_{11} + a_{11}^* - a_{13})(a_{22} + a_{22}^* - a_{24})-(a_{12} + a_{12}^* + a_{14} + a_{14}^*)^2} \, ,\label{ap2} \\
\sigma_{q_2,q_2}&=&\frac{2 (a_{11} + a_{11}^* - a_{13})}{4(a_{11} + a_{11}^* - a_{13})(a_{22} + a_{22}^* - a_{24})-(a_{12} + a_{12}^* + a_{14} + a_{14}^*)^2} \, ,\nonumber
\end{eqnarray}

By the use of the time derivatives of the convariance matrix of equation
(\ref{dsigma}), it can be demonstrated that the density matrix parameters
satisfy the following differential equations:
\begin{eqnarray}
\dot{a}_{11}&=&i \omega_{22} - 4 \omega_{12} a_{11} + 2 \omega_{23} a_{12} +i \omega_{11}
(-4 a_{11}^2 + a_{13}^2) + 2 i \omega_{13} (2 a_{11} a_{12} + a_{13} a_{14}) -  i \omega_{33} (a_{12}^2 - a_{14}^2) \, , \nonumber \\
 \dot{a}_{22}&=&i \omega_{44} + 2 \omega_{14} a_{12} - i \omega_{11} (a_{12}^2
 - a_{14}^{*2}) - 4 \omega_{34} a_{22} + 2 i \omega_{13} (2 a_{12} a_{22}
 + a_{14}^* a_{24}) +  i \omega_{33} (-4 a_{22}^2 +  a_{24}^2) \, , \nonumber \\
 \dot{a}_{12}&=&-2 i \omega_{24} + 4 \omega_{14} a_{11} - 2 \omega_{12} a_{12}
 - 2 \omega_{34} a_{12} - 2 i \omega_{11} (2 a_{11} a_{12} + a_{13} a_{14}^*) + 4 \omega_{23} a_{22} \nonumber \\
& & +  2 i \omega_{13} (a_{12}^2 - a_{14} a_{14}^* + 4 a_{11} a_{22} - a_{13} a_{24}) - 2 i \omega_{33}
(2 a_{12} a_{22} + a_{14} a_{24}) \, , \nonumber \\
 \dot{a}_{13}&=&-4 \omega_{12} a_{13} - 4 i \omega_{11} (a_{11} - a_{11}^*) a_{13}
  -  2 \omega_{23} (a_{14} + a_{14}^*) + 2 i \omega_{13} ((a_{12} - a_{12}^*) a_{13}  \label{parameters}    \\
 & &+ 2 a_{11}^* a_{14} - 2 a_{11} a_{14}^*)+ 2 i \omega_{33} (-a_{12}^* a_{14} + a_{12} a_{14}^*)  \, , \nonumber \\
\dot{a}_{14}&=&-2 \omega_{14} a_{13} - 2 \omega_{12} a_{14} - 2 \omega_{34} a_{14} - 2 i \omega_{11}
(a_{12}^* a_{13} + 2 a_{11} a_{14}) - 2 \omega_{23} a_{24}  \nonumber \\
& &+ 2 i \omega_{13} ((a_{12} - a_{12}^*) a_{14} + 2 a_{13} a_{22}^*
- 2 a_{11} a_{24}) + 2 i \omega_{33} (2 a_{14} a_{22}^* + a_{12} a_{24}) \, , \nonumber \\
 \dot{a}_{24}&=&-2 \omega_{14} (a_{14} + a_{14}^*) + 2 i \omega_{11} (a_{12} a_{14}
 - a_{12}^* a_{14}^*) - 4 \omega_{34} a_{24} +  4 i \omega_{33} (-a_{22} + a_{22}^*) a_{24} \nonumber \\
 & &+  2 i \omega_{13} (-2 a_{14} a_{22} + 2 a_{14}^* a_{22}^* + (a_{12} - a_{12}^*) a_{24}),
 \nonumber
\end{eqnarray}
which can be used to determine the time evolution of the Gaussian system at any time.

\section{B.~Matrix $\mathbf{M}$ for Bipartite System}
As discussed in section 4.2, the evolution of the covariance matrix of a bipartite system
can be written as the linear system of equations
\[
\mathbf{Mv}=0 \, ,
\]
where the vector containing the different covariances reads
\[
\tilde{v}=\left( \sigma_{p_1p_1}, \sigma_{p_1q_1},\sigma_{p_1p_2},\sigma_{p_1q_2},
\sigma_{q_1q_1},\sigma_{q_1p_2},\sigma_{q_1q_2},\sigma_{p_2p_2},\sigma_{p_2q_2},\sigma_{q_2 q_2} \right) \, ,
\]
and the matrix $\mathbf{M}$ is obtained by analyzing the derivatives
of the covariance matrix, this results in an expression for $\mathbf{M}$ given by

\begin{eqnarray}
{\scriptsize
\mathbf{M}=\left(
\begin{array}{cccccccccc}
 -4 \omega_{12} & -4 \omega_{22} & -4 \omega_{23} & -4 \omega_{24} & 0 & 0 & 0 & 0 & 0 & 0 \\
 2 \omega_{11} & 0 & 2 \omega_{13} & 2 \omega_{14} & -2 \omega_{22} & -2 \omega_{23} & -2 \omega_{24} & 0 & 0 & 0 \\
 -2 \omega_{14} & -2 \omega_{24} & -2 (\omega_{12}+\omega_{34}) & -2 \omega_{44} & 0 & -2 \omega_{22} & 0 & -2 \omega_{23} & -2 \omega_{24} & 0 \\
 2 \omega_{13} & 2 \omega_{23} & 2 \omega_{33} & 2 (\omega_{34}-\omega_{12}) & 0 & 0 & -2 \omega_{22} & 0 & -2 \omega_{23} & -2 \omega_{24} \\
 0 & 4 \omega_{11} & 0 & 0 & 4 \omega_{12} & 4 \omega_{13} & 4 \omega_{14} & 0 & 0 & 0 \\
 0 & -2 \omega_{14} & 2 \omega_{11} & 0 & -2 \omega_{24} & 2 (\omega_{12}-\omega_{34}) & -2 \omega_{44} & 2 \omega_{13} & 2 \omega_{14} & 0 \\
 0 & 2 \omega_{13} & 0 & 2 \omega_{11} & 2 \omega_{23} & 2 \omega_{33} & 2 (\omega_{12}+\omega_{34}) & 0 & 2 \omega_{13} & 2 \omega_{14} \\
 0 & 0 & -4 \omega_{14} & 0 & 0 & -4 \omega_{24} & 0 & -4 \omega_{34} & -4 \omega_{44} & 0 \\
 0 & 0 & 2 \omega_{13} & -2 \omega_{14} & 0 & 2 \omega_{23} & -2 \omega_{24} & 2 \omega_{33} & 0 & -2 \omega_{44} \\
 0 & 0 & 0 & 4 \omega_{13} & 0 & 0 & 4 \omega_{23} & 0 & 4 \omega_{33} & 4 \omega_{34} \\
\end{array}
\right)
}
\label{eme2}
\end{eqnarray}


\section*{References}

\begin{thebibliography}{99}
\bibitem{Khrennikov-book}
A. Khrennikov, {\it Contextual Approach to Quantum Formalism} (Springer:
Berlin/Heidelberg, Germany; New York, USA, 2009)

\bibitem{KhrenFP}
A. Khrennikov, Description of composite quantum systems by means of classical random
fields. Found. Phys., {\bf 40} 1051(2010). doi:10.1007/s10701-009-9392-8

\bibitem{khren1}
A. Khrennikov, G. Weihs, Preface of the special issue quantum foundations: Theory and experiment. Found. Phys.,  {\bf 42}, 721 (2012). doi:10.1007/s10701-012-9644-x

\bibitem{khren2}
G. M. D'Ariano, A. Khrennikov. Preface of the special issue quantum foundations: information approach. Philos. Trans. R. Soc. A Math. Phys. Engeen. Sci., {\bf 374}, 20150244 (2016).  doi:10.1098/rsta.2015.0244

\bibitem{khren3}
A. Khrennikov, and K. Svozil, Editorial: Quantum  probability and randomness. Entropy {\bf 21} 35 (2019). doi:3390/e21010035

\bibitem{gaussi}
X.-B. Wang, T. Hiroshima, A. Tomita, M. Hayashi. Quantum information with
Gaussian states, Physics Reports {\bf 448}, 1 (2007)
doi:10.1016/j.physrep.2007.04.005

\bibitem{adesso}
G. Adesso, S. Ragy, A. R. Lee. Continuous variable quantum information:
Gaussian states and beyond, Open Syst. Inf. Dyn. {\bf 21}, 1440001 (2014)
doi:10.1142/S1230161214400010

\bibitem{leuchs}
D. Bruss and G. Leuchs Eds. Quantum Information: From Foundations to
Quantum Technology Applications, (Wiley-VCH, Weinheim, 2019).

\bibitem{furasek}
J. Fiur\'asek, Gaussian Transformations and Distillation of Entangled Gaussian
States, Phys. Rev. Lett. {\bf 89} 137904 (2002).
doi:10.1103/PhysRevLett.89.137904

\bibitem{paris}
M. G. A. Paris, F. Illuminati, A. Serafini, and S. De Siena, Purity of
Gaussian states: Measurement schemes and time evolution in noisy channels,
Phys. Rev. A {\bf 68} 012314 (2003). doi:10.1103/PhysRevA.68.012314

\bibitem{serafini}
A. Serafini, F. Illuminati, and S. De Siena, Symplectic invariants, entropic
measures andcorrelations of Gaussian states, J. Phys. B: At. Mol. Opt. Phys.
{\bf 37} L21 (2004). doi:10.1088/0953-4075/37/2/L02

\bibitem{wolf}
M. M. Wolf, G. Giedke, and J. I. Cirac, Extremality of Gaussian Quantum
States, Phys. Rev. Lett. {\bf 96} 080502 (2006).
doi:10.1103/PhysRevLett.96.080502

\bibitem{cerf}
R. Garc\'ia-Patr\'on and N. J. Cerf, Unconditional Optimality of Gaussian
Attacks against Continuous Variable Quantum Key Distribution, Phys. Rev. Lett.
{\bf 97} 190503 (2006). doi:10.1103/PhysRevLett.97.190503

\bibitem{shapiro}
S.-H. Tan, B. I. Erkmen, V. Giovannetti, S. Guha, S. Lloyd, L. Maccone, S.
Pirandola, and J. H. Shapiro, Quantum Illumination with Gaussian States, Phys.
Rev. Lett. {\bf 101} 253601 (2008). doi:10.1103/PhysRevLett.101.253601

\bibitem{giorda}
P. Giorda, and M. G. A. Paris, Gaussian Quantum Discord, Phys. Rev. Lett {\bf
105} 020503 (2010). doi:10.1103/PhysRevLett.105.020503

\bibitem{arkhipov}
I. I. Arkhipov, J. Per\"ina Jr., J.   Svozilík, and A. Miranowicz,
Nonclassicality Invariant of General Two-Mode Gaussian States, Scien. Rep.,
{\bf 6} 26523 (2016). doi: 10.1038/srep26523

\bibitem{lami}
L. Lami, A. Serafini, and G. Adesso, Gaussian entanglement revisited, New J.
Phys. {\bf 20} 023030 (2018). doi:10.1088/1367-2630/aaa654

\bibitem{parrondo}
M. Mehboudi, J. M. R. Parrondo, and A. Ac\'in. Linear response theory for
quantum Gaussian processes, New J. Phys. {\bf 21} 083036 (2019).
doi:10.1088/1367-2630/ab30f4

\bibitem{jeong}
C. Oh, C. Lee, L. Banchi, S.-Y. Lee, C. Rockstuhl, and H. Jeong, Optimal
measurements for quantum fidelity between Gaussian states and its relevance to
quantum metrology, Phys. Rev. A  {\bf 100} 012323 (2019).
doi:10.1103/PhysRevA.100.012323

\bibitem{cruz1}
H. Cruz, D. Schuch, O. Casta\~nos, O. Rosas-Ortiz, Time-evolution of quantum
systems via a complex nonlinear Riccati equation. I. Conservative systems with
time-independent Hamiltonian, Annals of Phys. {\bf 360} 44 (2015).
doi:10.1016/j.aop.2015.05.001

\bibitem{cruz2}
H. Cruz, D. Schuch, O. Casta\~os, O. Rosas-Ortiz, Time-evolution of quantum
systems via a complex nonlinear Riccati equation. II. Dissipative systems,
Annals of Phys. {\bf  373} 609 (2016). doi:10.1016/j.aop.2016.07.029

\bibitem{dieter}
D. Schuch, Quantum theory from a nonlinear perspective: Riccati equations in
Fundamental physics, (Springer Nature, Cham, Switzerland, 2018).
doi:10.1007/978-3-31965594-9

\bibitem{dod_book}
V. V. Dodonov, {\it Coherent States and Their Generalizations for a Charged Particle in a Magnetic Field}, In: Antoine JP., Bagarello F., Gazeau JP. (eds) Coherent States and Their Applications (Springer Proceedings in Physics, vol 205. Springer, Cham, 2018) doi:10.1007/978-3-319-76732-1\_15

\bibitem{vale}
J. P. Valeriano and V. V. Dodonov, Non-monotonous behavior of the number variance, Mandel factor, invariant uncertainty product and purity for the quantum damped harmonic oscillator, Phys. Lett. A {\bf 384} 126370 (2020). doi:10.1016/j.physleta.2020.126370

\bibitem{Schroed26}
E.  Schr\"odinger, Quantisierung als Eigenwertproblem (Zweite Mitteilung). Ann.
Phys.  {\bf 384} 361, and 489 (1926). doi:10.1002/andp.19263840404

\bibitem{Landau}
L. Landau, Das D\"ampfungsproblem in der Wellenmechanik. Z. Phys. {\bf 45} 430 (1927).

\bibitem{vonNeumann}
J. von Neumann, Wahrscheinlichkeitstheoretischer Aufbau der Quantenmechanik.
G\"ott. Nach. 245 (1927).

\bibitem{kossa}
A. Kossakowski, On quantum statistical mechanics of non-Hamiltonian systems,
Rep. Math. Phys. {\bf 3} 247 (1972). doi:10.1016/0034-4877(72)90010-9

\bibitem{lind}
G. Lindblad, On the generators of quantum dynamical semigroups, Commun. Math.
Phys. {\bf 48}  119 (1976). doi:10.1007/BF01608499

\bibitem{gori}
V. Gorini, A. Kossakowski, E.C.G. Sudarshan, Completely positive semigroups of
N-level systems, J. Math. Phys. {\bf 17}  821 (1976). doi:10.1063/1.522979


\bibitem{vdod1}
V. V. Dodonov, O. V. Man'ko: Quantum damped oscillator in a magnetic field,
Physica A {\bf 130} 353 (1985). doi:10.1016/0378-4371(85)90111-6

\bibitem{vdod2}
V. V.Dodonov, O. V.Man'ko, and V. I.Man'ko: Quantum nonstationary oscillator: models and applications, J. Russ. Laser Res. {\bf 16}1 (1995). doi:10.1007/BF02581075

\bibitem{inv1}
V. V. Dodonov, I. A. Malkin, and V. I. Man'ko. J. Phys. A: Math. Gen. {\bf 8} L19 (1975). doi:10.1088/0305-4470/8/2/001

\bibitem{inv2}
V. V. Dodonov and V. I. Man'ko,   {\it Invariants and the evolution
of nonstationary quantum systems}, Proceedings of the Lebedev
Physical Institute, vol. 183, (Nova Science Publishers, New York,
1989).

\bibitem{Dodon-}
V. V. Dodonov, I. A. Malkin, and V. I. Man'ko, Integrals of the motion, Green
functions, and coherent states of dynamical systems, Int. J. Theor. Phys. {\bf
14} 3754 (1975)


\bibitem{sandulescu}
A. Sandulescu, H. Scutaru, and W. Scheid, Open quantum system of two coupled
harmonic oscillators for application in deep inelastic heavy ion collisions,
J. Phys. A: Math. Gen. {\bf 20} 2121 (1987). doi:10.1088/0305-4470/20/8/026

\bibitem{isar09}
A. Isar, Entanglement Generation and Evolution in Open Quantum Systems, Open
Sys. Inf. Dyn. {\bf 16}, 205 (2009). doi:10.1142/S1230161209000153

\bibitem{isar18}
A. Isar, Generation of quantum correlations in bipartite gaussian open quantum
systems, EPJ Web of Conferences {\bf 173} 01006 (2018).
doi:10.1051/epjconf/201817301006


\bibitem{Ramon-JRLR}
J. A. L\'opez-Sald\'ivar, A. Figueroa, O. Casta\~nos, R. L\'opez-Pe\~na, M. A.
Man'ko, and V. I. Man'ko, Evolution and entanglement of Gaussian states in the
parametric amplifier. J. Russ. Laser Res. {\bf 37} 23 (2016)
doi:10.1007/s10946-016-9543-2

\bibitem{Entropy-2-692-2018}
M.A. Man'ko and V.I. Man'ko, New Entropic Inequalities and Hidden Correlations
in Quantum Suprematism Picture of Qudit States. Entropy {\bf 20} 692 (2018). doi:10.3390/e20090692


\bibitem{Milestones}
M. A. Man'ko and V. I. Man'ko, From quantum carpets to quantum suprematism --the
probability representation of qudit states and hidden correlations. Phys.
Scr. {\bf 93} 084002 (2018). doi:10.1088/1402-4896/aacf24

\bibitem{Roberto}
R. Grimaudo, V. I. Man'ko, M. A. Man'ko and A. Messina. Dynamics of a harmonic
oscillator coupled with a Glauber amplifier. Phys. Scr. {\bf 95} 024004 (2020).
doi:10.1088/1402-4896/ab4305

\bibitem{PLA2020}
V.A. Andreev, M.A. Man'ko, V.I. Man'ko, Quantizer--dequantizer operators as
a tool for formulating the quantization procedure, Phys. Lett. A, in
press, Available online 26 February 2020.

\bibitem{Julio}
J. A. L\'opez-Sald\'ivar, General superposition states associated to the
rotational and inversion symmetries in the phase space. Phys. Scr. {\bf 95} 065206 (2020).
doi:10.1088/1402-4896/ab7feb

\bibitem{TombesiPLA}
S. Mancini, V. I.  Man'ko, P. Tombesi, Symplectic Tomography as Classical
Approach to Quantum Systems. Phys. Lett. A {\bf 213} 1 (1996). doi:10.1016/0375-9601(96)00107-7

\bibitem{Raymer1993}
D. T. Smithey,M. Beck, M. G. Raymer, and A. Faridani
Measurement of the Wigner distribution and the density matrix of a light mode
using optical homodyne tomography: Application to squeezed states and the
vacuum, Phys. Rev. Lett. {\bf 70} 1244 (1993). doi:10.1103/PhysRevLett.70.1244

\bibitem{TombesiFP}
S. Mancini, V. I. Man'ko, and P. Tombesi, Classical-like description of quantum
dynamics by means of symplectic tomography. Found. Phys. {\bf 27} 801 (1997).
doi:10.1007/BF02550342

\bibitem{KorennoyJRLR}
Y. A. Korennoy, V. I. Man'ko, Probability representation of the quantum
evolution and energy-level equations for optical tomograms. J Russ Laser Res
{\bf 32} 74 (2011). doi:10.1007/s10946-011-9191-5


\bibitem{dodcas}
V. V. Dodonov, Fifty years of dynamical Casimir effect. Physics, {\bf 2} 67
(2020). doi:10.3390/physics2010007

\bibitem{olgam}
O. V. Man'ko, Correlated squeezed states of a Josephson junction, J. Korean Phys. Soc. {\bf 27} 1 (1994)

\bibitem{dodosci}
V. V. Dodonov, O. V. Man'ko, and V. I. Man'ko, Correlated states in quantum electronics (resonant circuit), J.~Sov. Laser Res. {\bf 10}, 413 (1989). doi:10.1007/BF01120338.

\bibitem{IJQI-Turino}
M. A. Man'ko and V. I. Man'ko. Observables, interference phenomenon and Born's
rule in the probability representation of quantum mechanics, Int. J. Quantum
Inform. {\bf 18} 1941021 (2020).  doi:10.1142/S0219749919410211



\end{thebibliography}
\end{document}